\documentclass[orivec]{llncs}

\usepackage{amssymb}
\usepackage{amsmath}
\usepackage[ruled,vlined]{algorithm2e}
\usepackage{subfigure}
\usepackage{xcolor}
\usepackage[T1]{fontenc}
\usepackage{url}
\usepackage{array}
\usepackage{rotating}
\usepackage{floatrow}
\usepackage{tabularx}
\usepackage{tikz}
\usetikzlibrary{arrows,shapes,snakes,patterns}

\usepackage{todonotes}

\begin{document}

\title{Unifying Parsimonious Tree Reconciliation}

\author{Nicolas Wieseke \and Matthias Bernt \and Martin Middendorf}
\authorrunning{Wieseke et al.}

\institute{University of Leipzig, Faculty of Mathematics and Computer Science, Augustusplatz~10, 04109 Leipzig, Germany\\
\email{\{wieseke,bernt,middendorf\}@informatik.uni-leipzig.de}}

\maketitle

\begin{abstract}
Evolution is a process that is influenced by various environmental factors, e.g. the interactions between different species, genes, and  biogeographical properties. Hence, it is interesting to study the combined evolutionary history of multiple species, their genes, and the environment they live in. A common approach to address this research problem is to describe each individual evolution as a phylogenetic tree and construct a tree reconciliation which is parsimonious with respect to a given event model. Unfortunately, most of the previous approaches are designed only either for host-parasite systems, for gene tree/species tree reconciliation, or biogeography. Hence, a method is desirable, which addresses the general problem of mapping phylogenetic trees and covering all varieties of coevolving systems, including e.g., predator-prey and symbiotic relationships. To overcome this gap, we introduce a generalized cophylogenetic event model considering the combinatorial complete set of local coevolutionary events. We give a dynamic programming based heuristic for solving the maximum parsimony reconciliation problem in time $O(n^2)$, for two phylogenies each with at most $n$ leaves. Furthermore, we present an exact branch-and-bound algorithm which uses the results from the dynamic programming heuristic for discarding partial reconciliations. The approach has been implemented as a Java application which is freely available from
\url{http://pacosy.informatik.uni-leipzig.de/coresym}.
\end{abstract}

\section{Introduction}
\label{sec:introduction}
Tree reconciliation analysis is a powerful tool in phylogenetics and has a wide variety of applications. It is used in cladistic biogeography as well as for studying host-parasite coevolution and gene/species tree inference \cite{Page:98b}. A common principle for creating tree reconciliations is to use event-based maximum parsimony~\cite{Ronquist:90}. Therefore, coevolutionary events are defined together with a cost model for the events and  a reconciliation of the trees is sought that minimizes the overall costs.
Table \ref{tab:events} shows the common events used within the different types of applications.
Two closely related types of this problem can be distinguished: tree inference and tree embedding/tree mapping. In the first case an overarching tree is sought that embeds a given set of, possibly incongruent, trees. In the second case two (or more) trees are given and a mapping of those trees onto each other is sought. Although tree reconciliation refers to both types of problems, the scope of this paper lies on the latter case. 

Starting in 1979 Goodman et al. \cite{Goodman:79} introduced the problem of embedding gene trees into species trees. They presented a method to construct most parsimonious reconciliations, based on the evolutionary events \emph{gene duplication} and \emph{gene loss}. Since then, several algorithms for gene tree/species tree reconciliation have been developed, e.g., GeneTree \cite{Page:98}, SDI \cite{Zmasek:01}, Softparsmap \cite{Berglund:06}, Notung \cite{Chen:00,Vernot:08}, and Mowgli~\cite{Doyon:10}. These tools either consider additional events like \emph{lateral gene transfer} or \emph{incomplete lineage sorting}, or they are extended to suit unresolved phylogenetic trees. Recently, \cite{Szollosi:13} examines lateral gene transfer to and from species that are not represented in the phylogenetic tree, i.e. extincted or unsampled species. The respective event was called \emph{lateral transfer from the dead}.

A similar problem arose in the field of biogeography, where species trees and area cladograms have to be reconciled. Nelson and Platnick \cite{Nelson:81} were the first who proposed general assumptions for inferring area cladograms from species trees that have been implemented in the software COMPONENT 1.5 \cite{Page:88}. There are several other approaches for inferring area cladograms from species trees, based on association matrices, e.g., component compatibility analysis \cite{Zandee:87}, Brooks parsimony analysis \cite{Brooks:90}, and three area statement analysis \cite{Nelson:91}. But all of them are pattern-based approaches and do not produce reconciled trees. The dispersal-vicariance analysis \cite{Ronquist:97} implemented in the software DIVA is an event-based method considering \emph{vicariance}, \emph{duplication}, \emph{dispersal}, and \emph{extinction} events. It supports tree inference as well as tree embedding.

In 1988 Hafner and Nadler published a cophylogenetic analysis on pocket gophers and their chewing lice parasites \cite{Hafner:88}.
Methods designed for biogeographical \cite{Nelson:81} as well as gene tree/species tree problems \cite{Hendy:84} to examine \emph{cospeciation} events in host-parasite systems have been proposed in \cite{Hafner:90}. In the same year Ronquist and Nylin \cite{Ronquist:90} suggested the usage of \emph{colonisation}, \emph{exclusion}, and \emph{successive specialization} events with given weights relative to the probability of each event. Based on these events they developed a method for reconstructing the evolutionary history for two given phylogenetic trees and an association matrix describing the associations between extant species. 
This type of data set can be represented graphically by a so called tanglegram  \cite{Charleston:98}. In the same paper Charleston developed a data structure, called jungles, and a method to construct all optimal solutions for the reconciliation problem with events \emph{cospeciation}, \emph{duplication}, \emph{lineage sorting}, and \emph{host switch} under a general weighting scheme. The approach was implemented in the software TreeMap 2.0 and extended in Tarzan~\cite{Merkle:05} by the use of additional timing information. Jane \cite{Conow:10} and CoRe-PA~\cite{Merkle:10} use dynamic programming to efficiently compute reconciliations based on the same event model. The latter tool also considered unresolved phylogenies and presented a method for approximating event costs automatically. Additionally a first attempt was given to handle parasites infesting multiple hosts. The current version 4 of Jane supports multi-host parasites and \emph{failure to diverge} events as well. Furthermore, it is able to automatically resolve polytomies.

In \cite{Page:94} Page already pointed out the similarity between the three different types of reconciliation problems and gave a unified definition of reconciled trees. He presented the software COMPONENT 2.0, which he applied to data sets from all three types of problems.

Gene tree/species tree, biogeography, and host-parasite reconciliation problems have in common, that there exists an overarching tree into which the other trees have to be embedded, i.e., the gene trees have to be embedded into a species tree, the species tree into an area cladogram, and the parasite tree into a host tree. Until now there is no event model and respective algorithms which considers the general problem of two trees which also can be equitable mapped onto each other, as it is the case for, e.g., symbiotic systems or gene-gene interactions. 
%, or weaker kinds of dependency as for instance predator-prey relationships.
To fill this gap and to straighten up the event proliferation we introduce a combinatorial complete event model of local association patterns. However, a certain type of application may consider a subset of all possible event types only. Therefore, the event model can be utilized by defining a cost model with infinity costs for neglected events. Furthermore, we present a dynamic programming heuristic as well as an exact branch-and-bound algorithm to construct tree reconciliations under the new event model. In \cite{Ovadia:11} it was shown that even the special case of host-parasite tree reconciliation considering the events \emph{cospeciation}, \emph{sorting}, \emph{duplication}, and \emph{host switching} is NP-complete.
Therefore, heuristic or approximation algorithms might be the only chance to obtain solutions in reasonable time for large data sets.
However, for smaller data sets an computing an optimal solution might be feasible.

\begin{table}
\begin{tabularx}{\textwidth}{ >{\centering\arraybackslash}p{0.6in} | >{\centering\arraybackslash}X | >{\centering\arraybackslash}X | >{\centering\arraybackslash}X }
	                                 & biogeography
                                     & gene tree/species tree
                                     & host-parasite\\
    \hline
	\tikzstyle{vertexG}=[draw,circle,thick,fill=gray!50,minimum size=5pt]
\tikzstyle{vertexW}=[draw,circle,thick,fill=white,minimum size=5pt]
\tikzstyle{edge} = [draw,thick]
\tikzstyle{edge2} = [draw,densely dotted,thick]
\vspace{-5pt}
\begin{tikzpicture}[scale=0.5]
	\node[vertexG] (s) at (0,-1) {$$};
	\node[vertexW] (t) at (0.5,-1) {$$};
	\path[edge] (0,0) -- (s);
	\path[edge2] (0.5,0) -- (t);
	\path[edge] (s) -- (-0.75,-2);
	\path[edge] (s) -- (0.75,-2);
	\path[edge2] (t) -- (-0.25,-2);
	\path[edge2] (t) -- (1.25,-2);
\end{tikzpicture}
\vspace{-30pt}\newline   & vicariance \cite{Ronquist:11}, \linebreak allopatric \linebreak speciation \cite{Ronquist:97}
	                                 & speciation \linebreak (null event)
                                     & cospeciation \cite{Page:03}, \linebreak codivergence \cite{Charleston:06}, \linebreak successive \linebreak specialisation \cite{Ronquist:90}\\
	\hline
	\tikzstyle{vertexG}=[draw,circle,thick,fill=gray!50,minimum size=5pt]
\tikzstyle{vertexW}=[draw,circle,thick,fill=white,minimum size=5pt]
\tikzstyle{edge} = [draw,thick]
\tikzstyle{edge2} = [draw,densely dotted,thick]
\vspace{-5pt}
\begin{tikzpicture}[scale=0.5]
	\path[edge] (0,0) -- (0,-2);
	\path[edge2] (0.25,-1) -- (0.25,-2);
	\path[edge2] (0.75,-1) -- (0.75,-2);
	\node[vertexW] (t) at (0.5,-1) {$$};
	\path[edge2] (0.5,0) -- (t);
\end{tikzpicture}
\vspace{-30pt}\newline    & duplication \cite{Ronquist:11}, \linebreak sympatric \linebreak speciation \cite{Ronquist:97}
                                     & gene duplication \cite{Goodman:79}
                                     & duplication \cite{Page:03}, \linebreak independent speciation \cite{Page:03}\\
	\hline
	\tikzstyle{vertexG}=[draw,circle,thick,fill=gray!50,minimum size=5pt]
\tikzstyle{vertexW}=[draw,circle,thick,fill=white,minimum size=5pt]
\tikzstyle{edge} = [draw,thick]
\tikzstyle{edge2} = [draw,densely dotted,thick]
\vspace{-5pt}
\begin{tikzpicture}[scale=0.5]
	\node[vertexG] (s) at (0,-1) {$$};
	\path[edge] (0,0) -- (s);
	\path[edge2] (0.5,0) -- (0.5,-1) -- (1.25,-2);
	\path[edge] (s) -- (-0.75,-2);
	\path[edge] (s) -- (0.75,-2);
\end{tikzpicture}
\vspace{-30pt}\newline        & partial extinction \cite{Ronquist:11}
                                     & -
                                     & sorting \cite{Page:03}, \linebreak partial extinction \cite{Page:03}, \linebreak missing the boat \cite{Page:03}\\
	\tikzstyle{vertexG}=[draw,circle,thick,fill=gray!50,minimum size=5pt]
\tikzstyle{vertexW}=[draw,circle,thick,fill=white,minimum size=5pt]
\tikzstyle{edge} = [draw,thick]
\tikzstyle{edge2} = [draw,densely dotted,thick]
\vspace{-5pt}
\begin{tikzpicture}[scale=0.5]
	\node[vertexG] (s) at (0,-1) {$$};
	\node[vertexW] (t) at (0.5,-1) {$$};
	\path[edge] (0,0) -- (s);
	\path[edge2] (0.5,0) -- (t);
	\path[edge] (s) -- (-0.75,-2);
	\path[edge] (s) -- (0.75,-2);
	\path[edge2] (t) -- (-0,-1.75) -- node {$\times$} (-0,-1.75);
	\path[edge2] (t) -- (1.25,-2);
\end{tikzpicture}
\vspace{-30pt}\newline & -
	                                 & speciation and loss \cite{Doyon:10}
	                                 & -\\
	\hline
	\tikzstyle{vertexG}=[draw,circle,thick,fill=gray!50,minimum size=5pt]
\tikzstyle{vertexW}=[draw,circle,thick,fill=white,minimum size=5pt]
\tikzstyle{edge} = [draw,thick]
\tikzstyle{edge2} = [draw,densely dotted,thick]
\vspace{-5pt}
\begin{tikzpicture}[scale=0.5]
	\path[edge] (0,0) -- (0,-2);
	\path[edge2] (0.25,0) -- node[anchor=north] {$\times$} (0.25,-1);
\end{tikzpicture}
\vspace{-30pt}\newline     & complete extinction \cite{Ronquist:11}
	                                 & gene loss \cite{Goodman:79}
	                                 & extinction \cite{Page:03}\\
	\hline
	\tikzstyle{vertexG}=[draw,circle,thick,fill=gray!50,minimum size=5pt]
\tikzstyle{vertexW}=[draw,circle,thick,fill=white,minimum size=5pt]
\tikzstyle{edge} = [draw,thick]
\tikzstyle{edge2} = [draw,densely dotted,thick]
\vspace{-5pt}
\begin{tikzpicture}[scale=0.5]
	\path[edge] (0,0) -- (0,-2);
	\path[edge] (1.25,0) -- (1.25,-2);
	\node[vertexW] (t) at (0.25,-1) {$$};
	\path[edge2] (0.25,0) -- (t) -- (0.25,-2);
	\path[edge2] (t) -- (1,-1) -- (1,-2);
\end{tikzpicture}
\vspace{-30pt}\newline         & (partial) dispersal \cite{Ronquist:11}
	                                 & horizontal/lateral gene transfer \cite{Chen:00,Szollosi:13}, \linebreak direct transfer \cite{Szollosi:13}
	                                 & host switch \cite{Page:03}, \linebreak partial switch \cite{Page:03}\\
	\hline
	\tikzstyle{vertexG}=[draw,circle,thick,fill=gray!50,minimum size=5pt]
\tikzstyle{vertexW}=[draw,circle,thick,fill=white,minimum size=5pt]
\tikzstyle{edge} = [draw,thick]
\tikzstyle{edge2} = [draw,densely dotted,thick]
\vspace{-5pt}
\begin{tikzpicture}[scale=0.5]
	\path[edge] (0,0) -- (0,-2);
	\path[edge] (1.25,0) -- (1.25,-2);
	\path[edge2] (0.25,0) -- (0.25,-1) -- (1,-1) -- (1,-2);
\end{tikzpicture}
\vspace{-30pt}\newline & complete dispersal \cite{Ronquist:11}
	                                 & -
	                                 & complete switch \cite{Page:03}\\
	\tikzstyle{vertexG}=[draw,circle,thick,fill=gray!50,minimum size=5pt]
\tikzstyle{vertexW}=[draw,circle,thick,fill=white,minimum size=5pt]
\tikzstyle{edge} = [draw,thick]
\tikzstyle{edge2} = [draw,densely dotted,thick]
\vspace{-5pt}
\begin{tikzpicture}[scale=0.5]
	\path[edge] (0,0) -- (0,-2);
	\path[edge] (1.25,0) -- (1.25,-2);
	\node[vertexW] (t) at (0.25,-1) {$$};
	\path[edge2] (0.25,0) -- (t) -- (0.25,-1.75) -- node {$\times$} (0.25,-1.75);
	\path[edge2] (t) -- (1,-1) -- (1,-2);
\end{tikzpicture}
\vspace{-30pt}\newline   & -
	                                 & transfer and loss \cite{Doyon:10}
	                                 & -\\
	\hline
	\tikzstyle{vertexG}=[draw,circle,thick,fill=gray!50,minimum size=5pt]
\tikzstyle{vertexW}=[draw,circle,thick,fill=white,minimum size=5pt]
\tikzstyle{edge} = [draw,thick]
\tikzstyle{edge2} = [draw,densely dotted,thick]
\tikzstyle{edge3} = [draw,densely dotted,snake=coil,segment length=8.3,segment aspect=0,thick]
\vspace{-5pt}
\begin{tikzpicture}[scale=0.5]
	\path[edge] (0,0) -- (0,-2);
	\path[edge] (1.25,0) -- (1.25,-2);
	%\node[vertexW] (t1) at (2.25,-0.5) {$$};
	%\node[vertexW] (t2) at (2.25,-1.5) {$$};
	\path[edge2] (0.25,0) -- (0.25,-0.25) -- (2.25,-0.25);
	\path[edge3] (2.25,-0.25) -- (2.25,-1.75);
	\path[edge2] (2.25,-1.75) -- (1,-1.75) -- (1,-2);
\end{tikzpicture}
\vspace{-30pt}\newline          & -
	                                 & lateral transfer from the dead \cite{Szollosi:13}, \linebreak indirect transfer \cite{Szollosi:13}
	                                 & -\\
	\hline
	\tikzstyle{vertexG}=[draw,circle,thick,fill=gray!50,minimum size=5pt]
\tikzstyle{vertexW}=[draw,circle,thick,fill=white,minimum size=5pt]
\tikzstyle{edge} = [draw,thick]
\tikzstyle{edge2} = [draw,densely dotted,thick]
\vspace{-5pt}
\begin{tikzpicture}[scale=0.5]
	\path[edge] (0,0) -- (0,-2);
	\path[edge2] (0.25,0) -- (0.25,-1) -- (1,-1);
\end{tikzpicture}
\vspace{-30pt}\newline        & -
	                                 & -
	                                 & takeoff \cite{Merkle:10}, \linebreak exclusion \cite{Ronquist:90}\\
	\hline
	\tikzstyle{vertexG}=[draw,circle,thick,fill=gray!50,minimum size=5pt]
\tikzstyle{vertexW}=[draw,circle,thick,fill=white,minimum size=5pt]
\tikzstyle{edge} = [draw,thick]
\tikzstyle{edge2} = [draw,densely dotted,thick]
\vspace{-5pt}
\begin{tikzpicture}[scale=0.5]
	\path[edge] (1.25,0) -- (1.25,-2);
	\path[edge2] (0.25,-1) -- (1,-1) -- (1,-2);
\end{tikzpicture}
\vspace{-30pt}\newline        & -
	                                 & -
                                     & landing \cite{Merkle:10}, \linebreak colonisation \cite{Ronquist:90}\\
	\hline
	\tikzstyle{vertexG}=[draw,circle,thick,fill=gray!50,minimum size=5pt]
\tikzstyle{vertexW}=[draw,circle,thick,fill=white,minimum size=5pt]
\tikzstyle{edge} = [draw,thick]
\tikzstyle{edge2} = [draw,densely dotted,thick]
\vspace{-5pt}
\begin{tikzpicture}[scale=0.5]
	\node[vertexG] (s) at (0,-1) {$$};
	\path[edge] (0,0) -- (s);
	\path[edge2] (0.5,0) -- (0.5,-1);
	\path[edge] (s) -- (-0.75,-2);
	\path[edge] (s) -- (0.75,-2);
% 	\path[edge2] (0.5,-1) -- (-0.25,-1.75);
% 	\path[edge2] (0.5,-1) -- (1.25,-1.75);
% 	\path[edge2] (-0.25,-1.75) -- (1.25,-1.75);
	\draw[densely dotted, thick] (0.5,-1) -- (-0.25,-1.75) -- (1.25,-1.75) -- (0.5,-1);
\end{tikzpicture}
\vspace{-30pt}\newline            & -
	                                 & -
                                     & failure to speciate/diverge \cite{Page:03,Charleston:06}\\
\end{tabularx}
\caption{Coevolutionary events with the various names used within the different types of applications - biogeography (BG), gene/species tree (GST) and host-parasite coevolution (HPC). Solid lines represent the overarching tree (i.e., the area cladogram, species tree, and host tree, respectively), dotted lines represent the embedded tree (i.e., species tree, parasite tree, and gene tree, respectively).}
\label{tab:events}
\end{table}

\section{Basic Notations and Preliminaries}
\label{sec:notations}
For the following formal description we select the cophylogenetic reconciliation of two dependent sets of species as reference problem. Other kinds of tree reconciliation problems, e.g., of species, areas, or genes, are covered analogously. 
% For that reason, we use the notation \emph{species} synonymous for species, areas, or genes.

The evolution of a set of species is usually depicted as a \emph{phylogenetic tree}, which is a tree $T=(V_T,E_T)$ with node set $V_T$, edge set $E_T$, and leaf set $L_T\subset V_T$.
In the context of phylogenetic trees an internal node $u \in V_T \setminus L_T$ refers to a speciation of an ancestral species $u$ into subspecies. An edge $(u',u) \in E_T$ represents  the time span of the existence of a species $u$ from its emergence after the speciation of $u'$ until its own speciation or the present day. We refer to an edge $(u',u)$ as $e_u$ or simply $u$ if it is clear from the context.
If not stated otherwise we assume a phylogenetic tree being binary and rooted, i.e., each node has an outdegree of either two (internal node) or zero (leaf node) and there is exactly one node, the root $\rho_T$, with indegree zero whereas all other nodes $u \in V_T \setminus \rho_T$ have indegree one.
For each internal node $u \in V_T \setminus L_T$ we denote the children of $u$ as $u_i$ with $i \in \{1,2\}$.
For technical reasons we introduce an artificial root $\rho'$ and an edge $(\rho',\rho)$. In that way it is possible to refer to the time span of the existence of root species $\rho$ by $e_{\rho}$, or simply edge $\rho$.
We define a partial order $\preceq_T$ on $V_T$ such that $u' \preceq_T u$, if and only if $u'$ lies on the path from $\rho$ to $u$. In addition, $u' \prec_T u$ if and only if $u' \preceq_T u$ and $u' \neq u$. Node $u'$ is called an ancestor of $u$ and $u$ a descendant of $u'$, respectively.
%For a node $u \in V(T)$ $T_{|u}$ denotes the subtree of $T$ with root $u$.
Furthermore, we define the \emph{timing} $\tau$ of a tree as $\tau : V_T \rightarrow \mathbb{R}$ such that $\forall u',u \in V_T$ it holds that $u' \preceq_T u \Rightarrow \tau(u') \leq \tau(u)$ and $u' \prec_T u \Rightarrow \tau(u') < \tau(u)$, respectively. In the evolutionary context $\tau(u)$ represents the point in time at which the speciation of $u$ took place.

Let $(S,T,\phi)$ be a pair of rooted binary trees $S=(V_S,E_S)$ and $T=(V_T,E_T)$ together with a mapping $\phi(s,t) : V_S \times V_T \rightarrow [0,1]$ representing inter-species association strengths measured by a value between zero and one. The strength $\phi(s,t)$ can be interpreted as an a priori probability of two species $s$ and $t$ being associated. The given definition of $\phi$ is a generalization of the leaf-to-leaf associations $\varphi$ defined in \cite{Charleston:98}, extended by association strengths for extant and ancestral species. According to the notion of tanglegrams we refer to such a tuple $(S,T,\phi)$ as an \emph{X-tanglegram}. 

A \emph{cophylogenetic reconciliation} for a given X-tanglegram $(S,T,\phi)$ can be described as a set of associations $\mathcal{R} \subseteq E_S \times E_T$ between edges, with $\mathcal{R}$ being the reconciled interactions between extant as well as ancestral species. Depending on the type of application additional constrains on the set $\mathcal{R}$ are required for a reconciliation to be phylogenetically meaningful, e.g., timing constraints. These will be discussed later on.
A \emph{sub-reconciliation} $\mathcal{R}_{|s,t}$ is a subset of $\mathcal{R}$ such that $(e_{u},e_{v}) \in \mathcal{R}_{|s,t} \Leftrightarrow (e_{u},e_{v}) \in \mathcal{R}$ and $s \preceq_S u$ and $t \preceq_T v$, i.e., $\mathcal{R}_{|s,t}$ is a reconciliation of the subtrees of $S$ and $T$ rooted at nodes $s$ and $t$, respectively.

Note that in contrast to previous approaches we describe a reconciliation as a mapping between edges. In \cite{Merkle:05} a reconciliation was assumed to be a bijective mapping $m : \{V_T\} \rightarrow \{V_S \cup E_S\}$ from the nodes of the parasite tree to the nodes and edges of the host tree. 
% However, from a given edge mapping $\mathcal{R}$ the respective node/edge mapping $\mathcal{Q} \subseteq \{V_S \cup E_S\} \times \{V_T \cup E_T\}$ can be derived as follows.
% Assume $(e_s,e_t) \in \mathcal{R}$ and $i,j \in \{1,2\}$.
% If i) for all children $s_i$ of $s$ all $(e_{s_i},e_t) \notin \mathcal{R}$, then it holds that $\exists t_j : (e_s,e_{t_j}) \in \mathcal{R} \rightarrow (e_s,t) \in \mathcal{Q}$.
% Equivalently, if ii) for all children $t_j$ of $t$ all $(e_s,e_{t_j}) \notin \mathcal{R}$, then it holds that $\exists s_i : (e_{s_i},e_t) \in \mathcal{R} \rightarrow (s,e_t) \in \mathcal{Q}$.
% Furthermore, if i) and ii) hold, then $\exists s_i,t_j : (e_{s_i},e_{t_j}) \in \mathcal{R} \rightarrow (s,t) \in \mathcal{Q}$.
However, from a given mapping $m$ the respective edge-to-edge mapping $\mathcal{R}$ can be derived as follows.
Let $(t,t_i) \in E_T$ be an edge from $T$ and $s \in V_S$ be the node for which $m(t)=s$ or $m(t)=e_s$. Furthermore, let $s'' \in V_S$ be a node for which $m(t_i)=s''$ or $m(t_i)=e_{s''}$. Then it holds that $(e_s,e_t),(e_{s''},e_{t_i}) \in \mathcal{R}$.
If $s$ is an ancestor of $s''$ then for all nodes $s', s \prec_S s' \prec_S s''$ it holds that $\{(e_{s'},e_{t_i})\} \in \mathcal{R}$. Additionally, if $m(t)=e_s$ is an edge mapping then also $(e_s,e_{t_i}) \in \mathcal{R}$.

\section{Methods}
\label{sec:methods}
\subsection{Generalized Coevolutionary Event Model}
\label{sec:events}
Previous reconciliation approaches always consider a certain set of cophylogenetic events. These events are designed to suit to a certain type of application and some of them are combinations of other events, e.g., \emph{speciation and loss} or \emph{transfer and loss}.
In this section a generalized event model is presented that covers all possible local association patterns, i.e., all possible associations between coincident edges of two nodes. Hence, this event model can be applied to most of the applications. In the following the term \emph{species} will be interchangeably used for \emph{edge}. This is because associations between species have a one-to-one correspondence to pairs of associated edges in the reconciliation $\mathcal{R}$.

In this approach a cophylogenetic event is defined as a relation between the sets of coincident edges of two nodes $s \in V_S$ and $t \in V_T$. Formally, an event is described as a subset of $\{e_s,e_{s_1},e_{s_2}\} \times \{e_t,e_{t_1},e_{t_2}\}$. For a pair of species $(s,t)$ we define a binary variable $b_{(s,t)} \in \{0,1\}$, such that $b_{(s,t)}=1$ if species $s$ and $t$ are assumed to be associated, i.e., $(e_s,e_t) \in \mathcal{R}$, otherwise $b_{(s,t)}=0$.
Hence, an event can be described as an association tuple $\vec{b}_{s,t} = (b_{s,t},b_{s_1,t},b_{s_2,t},b_{s,t_1},b_{s,t_2},b_{s_1,t_1},b_{s_1,t_2},b_{s_2,t_1},b_{s_2,t_2})$ of length nine.

A cost model $\gamma: \{0,1\}^9 \rightarrow \mathbb{R}$ is defined specifying a cost value for each of the $2^9$ cophylogenetic events. Among the events there are some events which are isomorphic, i.e., they are identical when changing the child order of $s_1$ and $s_2$, respectively $t_1$ and $t_2$. For instance, the event $\{(e_s,e_t),(e_{s_1},e_t)\}$ is isomorphic to the event $\{(e_s,e_t),(e_{s_2},e_t)\}$.
Although it is not required, these isomorphic events should have the same cost value and events with no associations at all should be regarded as \emph{null event} and therefore get a value of zero.

\begin{figure}
\begin{scriptsize}
\begin{center}
\begin{minipage}[t]{0.24\textwidth}
\begin{center}
\tikzstyle{vertexG}=[draw,circle,thick,fill=gray!50,minimum size=15pt]
\tikzstyle{vertexW}=[draw,circle,thick,fill=white,minimum size=15pt]
\tikzstyle{edge} = [draw,thick,->]

\begin{tikzpicture}[scale=0.75]
	\node[vertexG] (s) at (0,-1) {$s$};
	\node[vertexW] (t) at (1,-1) {$t$};
	\path[edge] (0,0) -- node[anchor=east] {$e_s$} (s);
	\path[edge] (1,0) -- node[anchor=west] {$e_t$} (t);
	\path[edge] (s) -- (-0.5,-2);
	\path[edge] (s) -- (0,-2);
	\path[edge] (t) -- (1,-2);
	\path[edge] (t) -- (1.5,-2);
	\node[anchor=north] (s1) at (-0.5,-2) {$e_{s_1}$};
	\node[anchor=north] (s2) at (0,-2) {$e_{s_2}$};
	\node[anchor=north] (t1) at (1,-2) {$e_{t_1}$};
	\node[anchor=north] (t2) at (1.5,-2) {$e_{t_2}$};
\end{tikzpicture}
$\vec{b}=(000000000)$\\
Null event
\end{center}
\end{minipage}
\begin{minipage}[t]{0.24\textwidth}
\begin{center}
\tikzstyle{vertexG}=[draw,circle,thick,fill=gray!50,minimum size=15pt]
\tikzstyle{vertexW}=[draw,circle,thick,fill=white,minimum size=15pt]
\tikzstyle{edge} = [draw,thick,->]

\begin{tikzpicture}[scale=0.75]
	\draw[pattern=horizontal lines] (0,-1) -- (0.5,-1) -- (0.5,-2) -- (0,-2) -- (0,-1);
	\node[vertexG] (s) at (0,-1) {$s$};
	\node[vertexW] (t) at (0.5,-1) {$t$};
	\path[edge] (-0.5,0) -- node[anchor=east] {$e_s$} (s);
	\path[edge] (1,0) -- node[anchor=west] {$e_t$} (t);
	\path[edge] (s) -- (0,-2);
	\path[edge] (s) -- node[anchor=south] {$e_{s_2}$} (-1,-1);
	\path[edge] (t) -- (0.5,-2);
	\path[edge] (t) -- node[anchor=south] {$e_{t_2}$} (1.5,-1);
	\node[anchor=north] (s1) at (0,-2) {$e_{s_1}$};
	\node[anchor=north] (t1) at (0.5,-2) {$e_{t_1}$};
\end{tikzpicture}
$\vec{b}=(000001000)$\\
Double takeoff\\
on landing
\end{center}
\end{minipage}
\begin{minipage}[t]{0.24\textwidth}
\begin{center}
\tikzstyle{vertexG}=[draw,circle,thick,fill=gray!50,minimum size=15pt]
\tikzstyle{vertexW}=[draw,circle,thick,fill=white,minimum size=15pt]
\tikzstyle{edge} = [draw,thick,->]

\begin{tikzpicture}[scale=0.75]
	\draw[pattern=horizontal lines] (0,-1) -- (1,-1) -- (1,-2) -- (0,-2) -- (0,-1);
	%\draw[pattern=north west lines] (0,-1) -- (1,-1) -- (1,-2) -- (0,-2) -- (0,-1);
	%\draw[pattern=north east lines] (0,-1) -- (0.5,-1) -- (0.5,-2) -- (0,-2) -- (0,0);
	\path[edge] (0.5,-1) -- (0.5,-2);
	\path[edge] (1,-1) -- (1,-2);
	\node[vertexG] (s) at (0,-1) {$s$};
	\node[vertexW] (t) at (0.75,-1) {$t$};
	\path[edge] (-0.5,0) -- node[anchor=east] {$e_s$} (s);
	\path[edge] (1.25,0) -- node[anchor=west] {$e_t$} (t);
	\path[edge] (s) -- (0,-2);
	\path[edge] (s) -- node[anchor=south] {$e_{s_2}$} (-1,-1);
	\node[anchor=north] (s1) at (0,-2) {$e_{s_1}$};
	\node[anchor=north] (t1) at (0.5,-2) {$e_{t_1}$};
	\node[anchor=north] (t2) at (1,-2) {$e_{t_2}$};
\end{tikzpicture}
$\vec{b}=(000001100)$\\
S-takeoff + T-duplication\\
on landing
\end{center}
\end{minipage}
\begin{minipage}[t]{0.24\textwidth}
\begin{center}
\tikzstyle{vertexG}=[draw,circle,thick,fill=gray!50,minimum size=15pt]
\tikzstyle{vertexW}=[draw,circle,thick,fill=white,minimum size=15pt]
\tikzstyle{edge} = [draw,thick,->]

\begin{tikzpicture}[scale=0.75]
	\draw[pattern=horizontal lines] (-0.25,-1) -- (0.75,-1) -- (0.75,-2) -- (-0.25,-2) -- (-0.25,-1);
	%\draw[pattern=north east lines] (-0.25,-1) -- (0.75,-1) -- (0.75,-2) -- (-0.25,-2) -- (-0.25,-1);
	%\draw[pattern=north west lines] (0.25,-1) -- (0.75,-1) -- (0.75,-2) -- (0.25,-2) -- (0.25,-1);
	\path[edge] (-0.25,-1) -- (-0.25,-2);
	\path[edge] (0.25,-1) -- (0.25,-2);
	\node[vertexG] (s) at (0,-1) {$s$};
	\node[vertexW] (t) at (0.75,-1) {$t$};
	\path[edge] (-0.5,0) -- node[anchor=east] {$e_s$} (s);
	\path[edge] (1.25,0) -- node[anchor=west] {$e_t$} (t);
	\path[edge] (t) -- (0.75,-2);
	\path[edge] (t) -- node[anchor=south] {$e_{t_2}$} (1.5,-1);
	\node[anchor=north] (s1) at (-0.25,-2) {$e_{s_1}$};
	\node[anchor=north] (s2) at (0.25,-2) {$e_{s_2}$};
	\node[anchor=north] (s1) at (0.75,-2) {$e_{t_1}$};
\end{tikzpicture}
$\vec{b}=(000001010)$\\
T-takeoff + S-duplication
on landing
\end{center}
\end{minipage}
\begin{minipage}[t]{0.24\textwidth}
\begin{center}
\tikzstyle{vertexG}=[draw,circle,thick,fill=gray!50,minimum size=15pt]
\tikzstyle{vertexW}=[draw,circle,thick,fill=white,minimum size=15pt]
\tikzstyle{edge} = [draw,thick,->]

\begin{tikzpicture}[scale=0.75]
	\draw[pattern=north west lines] (0,-1) -- (0.5,-1) -- (-0.5,-2) -- (-1,-2) -- (0,-1);
	\draw[pattern=north east lines] (0,-1) -- (0.5,-1) -- (1.5,-2) -- (1,-2) -- (0,-1);
	\node[vertexG] (s) at (0,-1) {$s$};
	\node[vertexW] (t) at (0.5,-1) {$t$};
	\path[edge] (-0.5,0) -- node[anchor=east] {$e_s$} (s);
	\path[edge] (1,0) -- node[anchor=west] {$e_t$} (t);
	\path[edge] (s) -- (-1,-2);
	\path[edge] (s) -- (1,-2);
	\path[edge] (t) -- (-0.5,-2);
	\path[edge] (t) -- (1.5,-2);
	\node[anchor=north] (s1) at (-1,-2) {$e_{s_1}$};
	\node[anchor=north] (s2) at (1,-2) {$e_{s_2}$};
	\node[anchor=north] (t1) at (-0.5,-2) {$e_{t_1}$};
	\node[anchor=north] (t2) at (1.5,-2) {$e_{t_2}$};
\end{tikzpicture}
$\vec{b}=(000001001)$\\
Cospeciation\\
on landing
\end{center}
\end{minipage}
\begin{minipage}[t]{0.24\textwidth}
\begin{center}
\tikzstyle{vertexG}=[draw,circle,thick,fill=gray!50,minimum size=15pt]
\tikzstyle{vertexW}=[draw,circle,thick,fill=white,minimum size=15pt]
\tikzstyle{edge} = [draw,thick,->]

\begin{tikzpicture}[scale=0.75]
	\draw[fill=gray!50] (-0.25,-1.25) -- (0.5,-2) -- (-1,-2) -- (-0.25,-1.25);
	\draw[pattern=north west lines] (0,-1) -- (0.5,-1) -- (-0.5,-2) -- (-1,-2) -- (0,-1);
	\draw[pattern=north east lines] (0,-1) -- (0.5,-1) -- (1.5,-2) -- (0.5,-2) -- (-0.25,-1.25) -- (0,-1);
	\path[edge] (0,-1) -- node[anchor=east] {$e_{s_1}$} (-1,-2);
	\path[edge] (-0.25,-1.25) -- (0.5,-2);
	\draw[thick] (0.5,-2) -- (-1,-2);
	\path[edge] (0,-1) -- (1,-2);
	\path[edge] (0.5,-1) -- (-0.5,-2);
	\path[edge] (0.5,-1) -- node[anchor=west] {$e_{t_2}$} (1.5,-2);
	\node[vertexG] (s) at (0,-1) {$s$};
	\node[vertexW] (t) at (0.5,-1) {$t$};
	\path[edge] (-0.5,0) -- node[anchor=east] {$e_s$} (s);
	\path[edge] (1,0) -- node[anchor=west] {$e_t$} (t);
	\node[anchor=north] (s1) at (0.5,-2) {$e_{s_1}$};
	\node[anchor=north] (s2) at (1,-2) {$e_{s_2}$};
	\node[anchor=north] (t1) at (-0.5,-2) {$e_{t_1}$};
\end{tikzpicture}
$\vec{b}=(000001110)$\\
Partial cospeciation\\
on landing
\end{center}
\end{minipage}
\begin{minipage}[t]{0.24\textwidth}
\begin{center}
\tikzstyle{vertexG}=[draw,circle,thick,fill=gray!50,minimum size=15pt]
\tikzstyle{vertexW}=[draw,circle,thick,fill=white,minimum size=15pt]
\tikzstyle{edge} = [draw,thick,->]

\begin{tikzpicture}[scale=0.75]
	\draw[fill=lightgray!50] (0.75,-1.25) -- (0,-2) -- (1.5,-2) -- (0.75,-1.25);
	\draw[pattern=north east lines] (0,-1) -- (0.5,-1) -- (1.5,-2) -- (1,-2) -- (0,-1);
	\draw[pattern=north west lines] (0,-1) -- (0.5,-1) -- (0.75,-1.25) -- (0,-2) -- (-1,-2) -- (0,-1);
	\path[edge] (0,-1) -- node[anchor=east] {$e_{s_1}$} (-1,-2);
	\path[edge] (0,-1) -- (1,-2);
	\path[edge] (0.5,-1) -- (-0.5,-2);
	\path[edge] (0.75,-1.25) -- (0,-2);
	\path[edge] (0.5,-1) -- node[anchor=west] {$e_{t_2}$} (1.5,-2);
	\draw[thick] (0,-2) -- (1.5,-2);
	\node[vertexG] (s) at (0,-1) {$s$};
	\node[vertexW] (t) at (0.5,-1) {$t$};
	\path[edge] (-0.5,0) -- node[anchor=east] {$e_s$} (s);
	\path[edge] (1,0) -- node[anchor=west] {$e_t$} (t);
	\node[anchor=north] (s2) at (1,-2) {$e_{s_2}$};
	\node[anchor=north] (t1) at (-0.5,-2) {$e_{t_1}$};
	\node[anchor=north] (s1) at (0,-2) {$e_{t_2}$};
\end{tikzpicture}
$\vec{b}=(000000111)$\\
Partial cospeciation\\
on landing
\end{center}
\end{minipage}
\begin{minipage}[t]{0.24\textwidth}
\begin{center}
\tikzstyle{vertexG}=[draw,circle,thick,fill=gray!50,minimum size=15pt]
\tikzstyle{vertexW}=[draw,circle,thick,fill=white,minimum size=15pt]
\tikzstyle{edge} = [draw,thick,->]

\begin{tikzpicture}[scale=0.75]
	\draw[fill=lightgray!50] (0.75,-1.25) -- (0,-2) -- (1.5,-2) -- (0.75,-1.25);
	\draw[fill=gray!50] (-0.25,-1.25) -- (0.5,-2) -- (-1,-2) -- (-0.25,-1.25);
	%\draw[pattern=horizontal lines] (0,-1) -- (0.5,-1) -- (1.5,-2) -- (-1,-2) -- (0,-1);
	\draw[pattern=north west lines] (0,-1) -- (0.5,-1) -- (0.75,-1.25) -- (0,-2) -- (-1,-2) -- (0,-1);
	\draw[pattern=north east lines] (0,-1) -- (0.5,-1) -- (1.5,-2) -- (0.5,-2) -- (-0.25,-1.25) -- (0,-1);
	\path[edge] (0,-1) -- node[anchor=east] {$e_{s_1}$} (-1,-2);
	\path[edge] (-0.25,-1.25) -- (0.5,-2);
	\draw[thick] (0.5,-2) -- (-1,-2);
	\path[edge] (0,-1) -- (1,-2);
	\path[edge] (0.5,-1) -- (-0.5,-2);
	\path[edge] (0.75,-1.25) -- (0,-2);
	\path[edge] (0.5,-1) -- node[anchor=west] {$e_{t_2}$} (1.5,-2);
	\draw[thick] (0,-2) -- (1.5,-2);
	\node[vertexG] (s) at (0,-1) {$s$};
	\node[vertexW] (t) at (0.5,-1) {$t$};
	\path[edge] (-0.5,0) -- node[anchor=east] {$e_s$} (s);
	\path[edge] (1,0) -- node[anchor=west] {$e_t$} (t);
	\node[anchor=north] (s1) at (0.5,-2) {$e_{s_1}$};
	\node[anchor=north] (s2) at (1,-2) {$e_{s_2}$};
	\node[anchor=north] (t1) at (-0.5,-2) {$e_{t_1}$};
	\node[anchor=north] (s1) at (0,-2) {$e_{t_2}$};
\end{tikzpicture}
$\vec{b}=(000001111)$\\
Failure to cospeciate\\
on landing
\end{center}
\end{minipage}
\begin{minipage}[t]{0.24\textwidth}
\begin{center}
\tikzstyle{vertexG}=[draw,circle,thick,fill=gray!50,minimum size=15pt]
\tikzstyle{vertexW}=[draw,circle,thick,fill=white,minimum size=15pt]
\tikzstyle{edge} = [draw,thick,->]

\begin{tikzpicture}[scale=0.75]
	\draw[pattern=horizontal lines] (0,0) -- (0.5,0) -- (0.5,-1) -- (0,-1) -- (0,0);
	\node[vertexG] (s) at (0,-1) {$s$};
	\node[vertexW] (t) at (0.5,-1) {$t$};
	\path[edge] (0,0) -- node[anchor=east] {$e_s$} (s);
	\path[edge] (0.5,0) -- node[anchor=west] {$e_t$} (t);
	\path[edge] (s) -- (-0.75,-2);
	\path[edge] (s) -- (-0.25,-2);
	\path[edge] (t) -- (0.75,-2);
	\path[edge] (t) -- (1.25,-2);
	\node[anchor=north] (s1) at (-0.5,-2) {$e_{s_1}$};
	\node[anchor=north] (s2) at (0,-2) {$e_{s_2}$};
	\node[anchor=north] (t1) at (1,-2) {$e_{t_1}$};
	\node[anchor=north] (t2) at (1.5,-2) {$e_{t_2}$};
\end{tikzpicture}
$\vec{b}=(100000000)$\\
Association loss
\end{center}
\end{minipage}
\begin{minipage}[t]{0.24\textwidth}
\begin{center}
\tikzstyle{vertexG}=[draw,circle,thick,fill=gray!50,minimum size=15pt]
\tikzstyle{vertexW}=[draw,circle,thick,fill=white,minimum size=15pt]
\tikzstyle{edge} = [draw,thick,->]

\begin{tikzpicture}[scale=0.75]
	\draw[pattern=horizontal lines] (0,0) -- (0.5,0) -- (0.5,-2) -- (0,-2) -- (0,0);
	\node[vertexG] (s) at (0,-1) {$s$};
	\node[vertexW] (t) at (0.5,-1) {$t$};
	\path[edge] (0,0) -- node[anchor=east] {$e_s$} (s);
	\path[edge] (0.5,0) -- node[anchor=west] {$e_t$} (t);
	\path[edge] (s) -- (0,-2);
	\path[edge] (s) -- node[anchor=south] {$e_{s_2}$} (-1,-1);
	\path[edge] (t) -- (0.5,-2);
	\path[edge] (t) -- node[anchor=south] {$e_{t_2}$} (1.5,-1);
	\node[anchor=north] (s1) at (0,-2) {$e_{s_1}$};
	\node[anchor=north] (t1) at (0.5,-2) {$e_{t_1}$};
\end{tikzpicture}
$\vec{b}=(100001000)$\\
Double takeoff
\end{center}
\end{minipage}
\begin{minipage}[t]{0.24\textwidth}
\begin{center}
\tikzstyle{vertexG}=[draw,circle,thick,fill=gray!50,minimum size=15pt]
\tikzstyle{vertexW}=[draw,circle,thick,fill=white,minimum size=15pt]
\tikzstyle{edge} = [draw,thick,->]

\begin{tikzpicture}[scale=0.75]
	\draw[pattern=horizontal lines] (0,0) -- (0.75,0) -- (0.75,-1) -- (1,-1) -- (1,-2) -- (0,-2) -- (0,0);
	%\draw[pattern=horizontal lines] (0,0) -- (0.75,0) -- (0.75,-1) -- (0,-1) -- (0,0);
	%\draw[pattern=north west lines] (0,-1) -- (1,-1) -- (1,-2) -- (0,-2) -- (0,-1);
	%\draw[pattern=north east lines] (0,-1) -- (0.5,-1) -- (0.5,-2) -- (0,-2) -- (0,0);
	\path[edge] (0.5,-1) -- (0.5,-2);
	\path[edge] (1,-1) -- (1,-2);
	\node[vertexG] (s) at (0,-1) {$s$};
	\node[vertexW] (t) at (0.75,-1) {$t$};
	\path[edge] (0,0) -- node[anchor=east] {$e_s$} (s);
	\path[edge] (0.75,0) -- node[anchor=west] {$e_t$} (t);
	\path[edge] (s) -- (0,-2);
	\path[edge] (s) -- node[anchor=south] {$e_{s_2}$} (-1,-1);
	\node[anchor=north] (s1) at (0,-2) {$e_{s_1}$};
	\node[anchor=north] (t1) at (0.5,-2) {$e_{t_1}$};
	\node[anchor=north] (t2) at (1,-2) {$e_{t_2}$};
\end{tikzpicture}
$\vec{b}=(100001100)$\\
S-takeoff + T-duplication
\end{center}
\end{minipage}
\begin{minipage}[t]{0.24\textwidth}
\begin{center}
\tikzstyle{vertexG}=[draw,circle,thick,fill=gray!50,minimum size=15pt]
\tikzstyle{vertexW}=[draw,circle,thick,fill=white,minimum size=15pt]
\tikzstyle{edge} = [draw,thick,->]

\begin{tikzpicture}[scale=0.75]
	\draw[pattern=horizontal lines] (0,0) -- (0.75,0) -- (0.75,-2) -- (-0.25,-2) -- (-0.25,-1) -- (0,-1) -- (0,0);
	%\draw[pattern=horizontal lines] (0,0) -- (0.75,0) -- (0.75,-1) -- (0,-1) -- (0,0);
	%\draw[pattern=north east lines] (-0.25,-1) -- (0.75,-1) -- (0.75,-2) -- (-0.25,-2) -- (-0.25,-1);
	%\draw[pattern=north west lines] (0.25,-1) -- (0.75,-1) -- (0.75,-2) -- (0.25,-2) -- (0.25,-1);
	\path[edge] (-0.25,-1) -- (-0.25,-2);
	\path[edge] (0.25,-1) -- (0.25,-2);
	\node[vertexG] (s) at (0,-1) {$s$};
	\node[vertexW] (t) at (0.75,-1) {$t$};
	\path[edge] (0,0) -- node[anchor=east] {$e_s$} (s);
	\path[edge] (0.75,0) -- node[anchor=west] {$e_t$} (t);
	\path[edge] (t) -- (0.75,-2);
	\path[edge] (t) -- node[anchor=south] {$e_{t_2}$} (1.5,-1);
	\node[anchor=north] (s1) at (-0.25,-2) {$e_{s_1}$};
	\node[anchor=north] (s2) at (0.25,-2) {$e_{s_2}$};
	\node[anchor=north] (s1) at (0.75,-2) {$e_{t_1}$};
\end{tikzpicture}
$\vec{b}=(100001010)$\\
T-takeoff + S-duplication
\end{center}
\end{minipage}
\begin{minipage}[t]{0.24\textwidth}
\begin{center}
\tikzstyle{vertexG}=[draw,circle,thick,fill=gray!50,minimum size=15pt]
\tikzstyle{vertexW}=[draw,circle,thick,fill=white,minimum size=15pt]
\tikzstyle{edge} = [draw,thick,->]

\begin{tikzpicture}[scale=0.75]
	\draw[pattern=horizontal lines] (0,0) -- (0.5,0) -- (0.5,-1) -- (0,-1) -- (0,0);
	\draw[pattern=north west lines] (0,-1) -- (0.5,-1) -- (-0.5,-2) -- (-1,-2) -- (0,-1);
	\draw[pattern=north east lines] (0,-1) -- (0.5,-1) -- (1.5,-2) -- (1,-2) -- (0,-1);
	\node[vertexG] (s) at (0,-1) {$s$};
	\node[vertexW] (t) at (0.5,-1) {$t$};
	\path[edge] (0,0) -- node[anchor=east] {$e_s$} (s);
	\path[edge] (0.5,0) -- node[anchor=west] {$e_t$} (t);
	\path[edge] (s) -- (-1,-2);
	\path[edge] (s) -- (1,-2);
	\path[edge] (t) -- (-0.5,-2);
	\path[edge] (t) -- (1.5,-2);
	\node[anchor=north] (s1) at (-1,-2) {$e_{s_1}$};
	\node[anchor=north] (s2) at (1,-2) {$e_{s_2}$};
	\node[anchor=north] (t1) at (-0.5,-2) {$e_{t_1}$};
	\node[anchor=north] (t2) at (1.5,-2) {$e_{t_2}$};
\end{tikzpicture}
$\vec{b}=(100001001)$\\
Cospeciation
\end{center}
\end{minipage}
\begin{minipage}[t]{0.24\textwidth}
\begin{center}
\tikzstyle{vertexG}=[draw,circle,thick,fill=gray!50,minimum size=15pt]
\tikzstyle{vertexW}=[draw,circle,thick,fill=white,minimum size=15pt]
\tikzstyle{edge} = [draw,thick,->]

\begin{tikzpicture}[scale=0.75]
	\draw[pattern=horizontal lines] (0,0) -- (0.5,0) -- (0.5,-1) -- (0,-1) -- (0,0);
	\draw[fill=gray!50] (-0.25,-1.25) -- (0.5,-2) -- (-1,-2) -- (-0.25,-1.25);
	\draw[pattern=north west lines] (0,-1) -- (0.5,-1) -- (-0.5,-2) -- (-1,-2) -- (0,-1);
	\draw[pattern=north east lines] (0,-1) -- (0.5,-1) -- (1.5,-2) -- (0.5,-2) -- (-0.25,-1.25) -- (0,-1);
	\path[edge] (0,-1) -- node[anchor=east] {$e_{s_1}$} (-1,-2);
	\path[edge] (-0.25,-1.25) -- (0.5,-2);
	\draw[thick] (0.5,-2) -- (-1,-2);
	\path[edge] (0,-1) -- (1,-2);
	\path[edge] (0.5,-1) -- (-0.5,-2);
	\path[edge] (0.5,-1) -- node[anchor=west] {$e_{t_2}$} (1.5,-2);
	\node[vertexG] (s) at (0,-1) {$s$};
	\node[vertexW] (t) at (0.5,-1) {$t$};
	\path[edge] (0,0) -- node[anchor=east] {$e_s$} (s);
	\path[edge] (0.5,0) -- node[anchor=west] {$e_t$} (t);
	\node[anchor=north] (s1) at (0.5,-2) {$e_{s_1}$};
	\node[anchor=north] (s2) at (1,-2) {$e_{s_2}$};
	\node[anchor=north] (t1) at (-0.5,-2) {$e_{t_1}$};
\end{tikzpicture}
$\vec{b}=(100001110)$\\
Partial cospeciation
\end{center}
\end{minipage}
\begin{minipage}[t]{0.24\textwidth}
\begin{center}
\tikzstyle{vertexG}=[draw,circle,thick,fill=gray!50,minimum size=15pt]
\tikzstyle{vertexW}=[draw,circle,thick,fill=white,minimum size=15pt]
\tikzstyle{edge} = [draw,thick,->]

\begin{tikzpicture}[scale=0.75]
	\draw[pattern=horizontal lines] (0,0) -- (0.5,0) -- (0.5,-1) -- (0,-1) -- (0,0);
	\draw[fill=lightgray!50] (0.75,-1.25) -- (0,-2) -- (1.5,-2) -- (0.75,-1.25);
	\draw[pattern=north east lines] (0,-1) -- (0.5,-1) -- (1.5,-2) -- (1,-2) -- (0,-1);
	\draw[pattern=north west lines] (0,-1) -- (0.5,-1) -- (0.75,-1.25) -- (0,-2) -- (-1,-2) -- (0,-1);
	\path[edge] (0,-1) -- node[anchor=east] {$e_{s_1}$} (-1,-2);
	\path[edge] (0,-1) -- (1,-2);
	\path[edge] (0.5,-1) -- (-0.5,-2);
	\path[edge] (0.75,-1.25) -- (0,-2);
	\path[edge] (0.5,-1) -- node[anchor=west] {$e_{t_2}$} (1.5,-2);
	\draw[thick] (0,-2) -- (1.5,-2);
	\node[vertexG] (s) at (0,-1) {$s$};
	\node[vertexW] (t) at (0.5,-1) {$t$};
	\path[edge] (0,0) -- node[anchor=east] {$e_s$} (s);
	\path[edge] (0.5,0) -- node[anchor=west] {$e_t$} (t);
	\node[anchor=north] (s2) at (1,-2) {$e_{s_2}$};
	\node[anchor=north] (t1) at (-0.5,-2) {$e_{t_1}$};
	\node[anchor=north] (s1) at (0,-2) {$e_{t_2}$};
\end{tikzpicture}
$\vec{b}=(100000111)$\\
Partial cospeciation
\end{center}
\end{minipage}
\begin{minipage}[t]{0.24\textwidth}
\begin{center}
\tikzstyle{vertexG}=[draw,circle,thick,fill=gray!50,minimum size=15pt]
\tikzstyle{vertexW}=[draw,circle,thick,fill=white,minimum size=15pt]
\tikzstyle{edge} = [draw,thick,->]

\begin{tikzpicture}[scale=0.75]
	\draw[pattern=horizontal lines] (0,0) -- (0.5,0) -- (0.5,-1) -- (0,-1) -- (0,0);
	\draw[fill=lightgray!50] (0.75,-1.25) -- (0,-2) -- (1.5,-2) -- (0.75,-1.25);
	\draw[fill=gray!50] (-0.25,-1.25) -- (0.5,-2) -- (-1,-2) -- (-0.25,-1.25);
	%\draw[pattern=horizontal lines] (0,-1) -- (0.5,-1) -- (1.5,-2) -- (-1,-2) -- (0,-1);
	\draw[pattern=north west lines] (0,-1) -- (0.5,-1) -- (0.75,-1.25) -- (0,-2) -- (-1,-2) -- (0,-1);
	\draw[pattern=north east lines] (0,-1) -- (0.5,-1) -- (1.5,-2) -- (0.5,-2) -- (-0.25,-1.25) -- (0,-1);
	\path[edge] (0,-1) -- node[anchor=east] {$e_{s_1}$} (-1,-2);
	\path[edge] (-0.25,-1.25) -- (0.5,-2);
	\draw[thick] (0.5,-2) -- (-1,-2);
	\path[edge] (0,-1) -- (1,-2);
	\path[edge] (0.5,-1) -- (-0.5,-2);
	\path[edge] (0.75,-1.25) -- (0,-2);
	\path[edge] (0.5,-1) -- node[anchor=west] {$e_{t_2}$} (1.5,-2);
	\draw[thick] (0,-2) -- (1.5,-2);
	\node[vertexG] (s) at (0,-1) {$s$};
	\node[vertexW] (t) at (0.5,-1) {$t$};
	\path[edge] (0,0) -- node[anchor=east] {$e_s$} (s);
	\path[edge] (0.5,0) -- node[anchor=west] {$e_t$} (t);
	\node[anchor=north] (s1) at (0.5,-2) {$e_{s_1}$};
	\node[anchor=north] (s2) at (1,-2) {$e_{s_2}$};
	\node[anchor=north] (t1) at (-0.5,-2) {$e_{t_1}$};
	\node[anchor=north] (s1) at (0,-2) {$e_{t_2}$};
\end{tikzpicture}
$\vec{b}=(100001111)$\\
Failure to cospeciate
\end{center}
\end{minipage}

\end{center}
\end{scriptsize}
\caption{Coevolutionary events for case iii), i.e., speciation of $s$ and $t$ occurring simultaneously.}
\label{fig:eqEvents}
\end{figure}
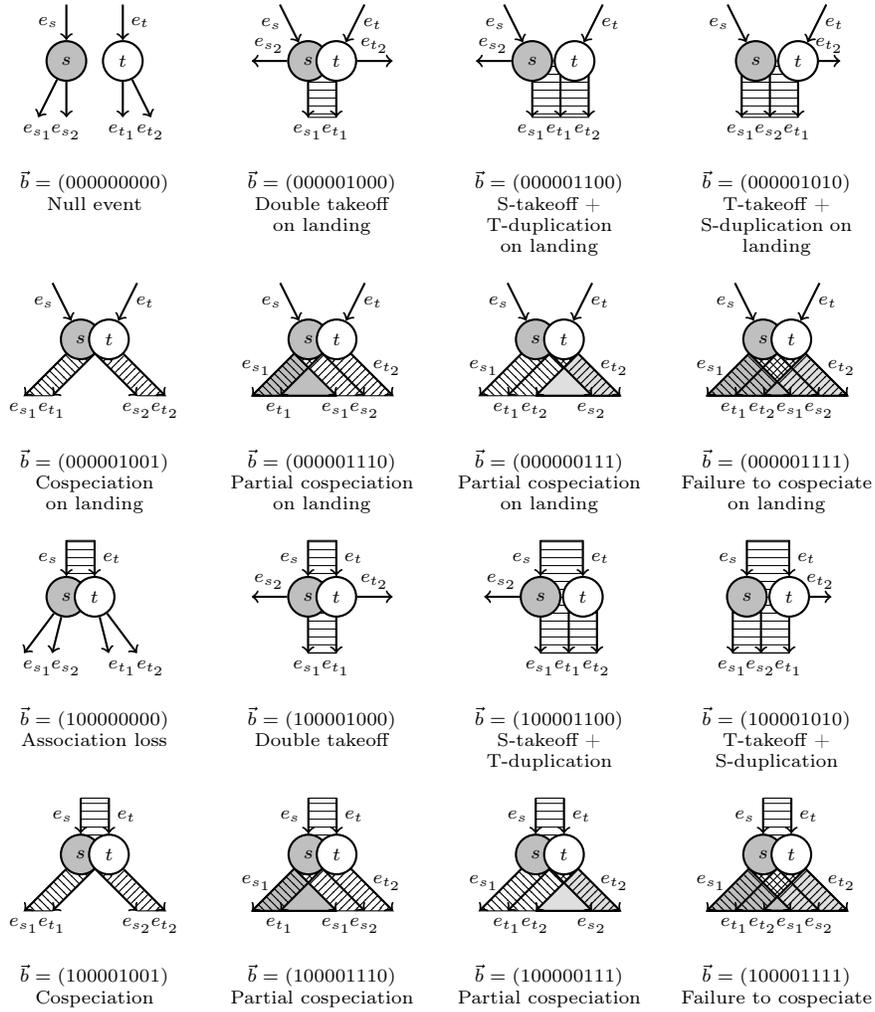

Not all events which can be modeled this way are phylogenetically meaningful. If, for example, species $s$ is associated with a child $t_j$ of species $t$ and $t$ with a child species $s_i$ of $s$, this would immediately result in an inconsistency as the speciation of $s$ had to occur before the speciation of $t$ and vice versa. Therefore, we distinguish between three types of events: i) events with $\tau(s)<\tau(t)$ (denoted as ``$<$'' events), ii) events with $\tau(s)>\tau(t)$ (``$>$'' events) and iii) events with $\tau(s)=\tau(t)$ (``$=$'' events). In the first case there must not be an association between $s$ and the descendant species $t_i$. The second case is equivalently but with no associations between $t$ and $s_j$. In the last case none of the species $s$, respectively $t$, is associated with the descendant species $t_j$, respectively $s_i$. The events for case i) and case ii) are depicted in Figure \ref{fig:legeEvents}. Case ii) is symmetric to the first case with exchanged roles of $s$ and $t$. Case iii) is shown in Figure \ref{fig:eqEvents}.

\begin{figure}
\begin{scriptsize}
\begin{center}
\begin{minipage}[t]{0.24\textwidth}
\begin{center}
\tikzstyle{vertexG}=[draw,circle,thick,fill=gray!50,minimum size=15pt]
\tikzstyle{vertexW}=[draw,circle,thick,fill=white,minimum size=15pt]
\tikzstyle{edge} = [draw,thick,->]

\begin{tikzpicture}[scale=0.75]
	\node[vertexG] (s) at (0,-1) {$s$};
	\node[vertexW] (t) at (1.25,-2) {$t$};
	\path[edge] (1.25,0) -- node[anchor=west] {$e_t$} (t);
	\path[edge] (0,0) -- node[anchor=east] {$e_s$} (s);
	\path[edge] (s) -- (-0.5,-2);
	\path[edge] (s) -- (0.5,-2);
	\node[anchor=north] (s1) at (-0.5,-2) {$e_{s_1}$};
	\node[anchor=north] (s2) at (0.5,-2) {$e_{s_2}$};
\end{tikzpicture}
$\vec{b}=(00000****)$\\
Null event
\end{center}
\end{minipage}
\begin{minipage}[t]{0.24\textwidth}
\begin{center}
\tikzstyle{vertexG}=[draw,circle,thick,fill=gray!50,minimum size=15pt]
\tikzstyle{vertexW}=[draw,circle,thick,fill=white,minimum size=15pt]
\tikzstyle{edge} = [draw,thick,->]

\begin{tikzpicture}[scale=0.75]
	\draw[pattern=north west lines] (-0.25,-1.25) -- (0,-1.5) -- (-0.5,-2) -- (-1,-2) -- (-0.25,-1.25);
	\node[vertexG] (s) at (0,-1) {$s$};
	\node[vertexW] (t) at (-0.5,-2) {$t$};
	\path[edge] (1,-1.25) -- node[anchor=south,pos=0.25] {$e_t$} (0,-1.5) -- (t);
	\path[edge] (0,0) -- node[anchor=east] {$e_s$} (s);
	\path[edge] (s) -- (-1,-2);
	\path[edge] (s) --(1,-2);
 	\node[anchor=north] (s1) at (-1,-2) {$e_{s_1}$};
	\node[anchor=north] (s2) at (1,-2) {$e_{s_2}$};
\end{tikzpicture}
$\vec{b}=(01000****)$\\
Landing
\end{center}
\end{minipage}
\begin{minipage}[t]{0.24\textwidth}
\begin{center}
\tikzstyle{vertexG}=[draw,circle,thick,fill=gray!50,minimum size=15pt]
\tikzstyle{vertexW}=[draw,circle,thick,fill=white,minimum size=15pt]
\tikzstyle{edge} = [draw,thick,->]

\begin{tikzpicture}[scale=0.75]
	\draw[pattern=north east lines] (0.75,-1.75) -- (1,-1.5) -- (1.5,-2) -- (1,-2) -- (0.75,-1.75);
	\node[vertexG] (s) at (0,-1) {$s$};
	\node[vertexW] (t) at (1.5,-2) {$t$};
	\path[edge] (1.75,-1.25) -- node[anchor=south] {$e_t$} (1,-1.5) -- (t);
	\path[edge] (0,0) -- node[anchor=east] {$e_s$} (s);
	\path[edge] (s) -- (-1,-2);
	\path[edge] (s) -- (1,-2);
	\node[anchor=north] (s1) at (-1,-2) {$e_{s_1}$};
	\node[anchor=north] (s2) at (1,-2) {$e_{s_2}$};
\end{tikzpicture}
$\vec{b}=(00100****)$\\
Landing
\end{center}
\end{minipage}
\begin{minipage}[t]{0.24\textwidth}
\begin{center}
\tikzstyle{vertexG}=[draw,circle,thick,fill=gray!50,minimum size=15pt]
\tikzstyle{vertexW}=[draw,ellipse,thick,fill=white,minimum width=30pt]
\tikzstyle{edge} = [draw,thick,->]

\begin{tikzpicture}[scale=0.75]
	\draw[fill=lightgray!50] (0,-1.5) -- (-0.5,-2) -- (0.5,-2) -- (0,-1.5);
	\draw[pattern=north west lines] (-0.25,-1.25) -- (0,-1.5) -- (-0.5,-2) -- (-1,-2) -- (-0.25,-1.25);
	\draw[pattern=north east lines] (0,-1.5) -- (0.25,-1.25) -- (1,-2) -- (0.5,-2) -- (0,-1.5);
	\node[vertexG] (s) at (0,-1) {$s$};
	\draw[thick] (1,-1.25) -- node[anchor=south,pos=0.25] {$e_t$} (0,-1.5);
	\path[edge] (0,-1.5) -- (-0.4,-1.9);
	\path[edge] (0,-1.5) -- (0.4,-1.9);
	\node[vertexW] (t) at (0,-2.1) {$t$};
	\path[edge] (0,0) -- node[anchor=east] {$e_s$} (s);
	\path[edge] (s) -- (-1,-2);
	\path[edge] (s) -- (1,-2);
	\node[anchor=north] (s1) at (-1,-2) {$e_{s_1}$};
	\node[anchor=north] (s2) at (1,-2) {$e_{s_2}$};
\end{tikzpicture}
$\vec{b}=(01100****)$\\
Double landing
\end{center}
\end{minipage}
\begin{minipage}[t]{0.24\textwidth}
\begin{center}
\tikzstyle{vertexG}=[draw,circle,thick,fill=gray!50,minimum size=15pt]
\tikzstyle{vertexW}=[draw,circle,thick,fill=white,minimum size=15pt]
\tikzstyle{edge} = [draw,thick,->]

\begin{tikzpicture}[scale=0.75]
	\draw[pattern=horizontal lines] (0,0) -- (0.5,0) -- (0.5,-0.5) -- (0,-0.5) -- (0,0);
	\node[vertexG] (s) at (0,-1) {$s$};
	\path[edge] (0.5,0) -- node[anchor=west] {$e_t$} (0.5,-0.5);
	\node (x) at (0.5,-0.5) {$\times$};
	\path[edge] (0,0) -- node[anchor=east] {$e_s$} (s);
	\path[edge] (s) -- (-1,-2);
	\path[edge] (s) -- (1,-2);
	\node[anchor=north] (s1) at (-1,-2) {$e_{s_1}$};
	\node[anchor=north] (s2) at (1,-2) {$e_{s_2}$};
\end{tikzpicture}
$\vec{b}=(100000000)$\\
Extinction / Association loss 
\end{center}
\end{minipage}
\begin{minipage}[t]{0.24\textwidth}
\begin{center}
\tikzstyle{vertexG}=[draw,circle,thick,fill=gray!50,minimum size=15pt]
\tikzstyle{vertexW}=[draw,circle,thick,fill=white,minimum size=15pt]
\tikzstyle{edge} = [draw,thick,->]

\begin{tikzpicture}[scale=0.75]
	\draw[pattern=horizontal lines] (0,0) -- (0.5,0) -- (0.5,-1) -- (-0.5,-2) -- (-1,-2) -- (0,-1) -- (0,0);
	\node[vertexG] (s) at (0,-1) {$s$};
	\node[vertexW] (t) at (-0.5,-2) {$t$};
	\path[edge] (0.5,0) -- node[anchor=west] {$e_t$} (0.5,-1) -- (t);
	\path[edge] (0,0) -- node[anchor=east] {$e_s$} (s);
	\path[edge] (s) -- (-1,-2);
	\path[edge] (s) -- (1,-2);
	\node[anchor=north] (s1) at (-1,-2) {$e_{s_1}$};
	\node[anchor=north] (s2) at (1,-2) {$e_{s_2}$};
\end{tikzpicture}
$\vec{b}=(11000****)$\\
T-sorting (S-takeoff)
\end{center}
\end{minipage}
\begin{minipage}[t]{0.24\textwidth}
\begin{center}
\tikzstyle{vertexG}=[draw,circle,thick,fill=gray!50,minimum size=15pt]
\tikzstyle{vertexW}=[draw,circle,thick,fill=white,minimum size=15pt]
\tikzstyle{edge} = [draw,thick,->]

\begin{tikzpicture}[scale=0.75]
	\draw[pattern=horizontal lines] (0,0) -- (0.5,0) -- (0.5,-1) -- (1.5,-2) -- (1,-2) -- (0,-1) -- (0,0);
	\node[vertexG] (s) at (0,-1) {$s$};
	\node[vertexW] (t) at (1.5,-2) {$t$};
	\path[edge] (0.5,0) -- node[anchor=west] {$e_t$} (0.5,-1) -- (t);
	\path[edge] (0,0) -- node[anchor=east] {$e_s$} (s);
	\path[edge] (s) -- (-1,-2);
	\path[edge] (s) -- (1,-2);
	\node[anchor=north] (s1) at (-1,-2) {$e_{s_1}$};
	\node[anchor=north] (s2) at (1,-2) {$e_{s_2}$};
\end{tikzpicture}
$\vec{b}=(10100****)$\\
T-sorting (S-takeoff)
\end{center}
\end{minipage}
\begin{minipage}[t]{0.24\textwidth}
\begin{center}
\tikzstyle{vertexG}=[draw,circle,thick,fill=gray!50,minimum size=15pt]
\tikzstyle{vertexW}=[draw,circle,thick,fill=white,minimum size=15pt]
\tikzstyle{edge} = [draw,thick,->]

\begin{tikzpicture}[scale=0.75]
	\draw[pattern=horizontal lines] (0,0) -- (0.75,0) -- (0.75,-2) -- (-0.25,-2) -- (-0.25,-1) -- (0,-1) -- (0,0);
	%\draw[pattern=horizontal lines] (0,0) -- (0.75,0) -- (0.75,-1) -- (0,-1) -- (0,0);
	%\draw[pattern=north east lines] (-0.25,-1) -- (0.75,-1) -- (0.75,-2) -- (-0.25,-2) -- (-0.25,-1);
	%\draw[pattern=north west lines] (0.25,-1) -- (0.75,-1) -- (0.75,-2) -- (0.25,-2) -- (0.25,-1);
	\path[edge] (0,0) -- node[anchor=east] {$e_s$} (s);
	\path[edge] (-0.25,-1) -- (-0.25,-2);
	\path[edge] (0.25,-1) -- (0.25,-2);
	\node[vertexG] (s) at (0,-1) {$s$};
	\node[vertexW] (t) at (0.75,-2) {$t$};
	\path[edge] (0.75,0) -- node[anchor=west] {$e_t$} (t);
 	\node[anchor=north] (s2) at (-0.25,-2) {$e_{s_1}$};
 	\node[anchor=north] (s1) at (0.25,-2) {$e_{s_2}$};
\end{tikzpicture}
$\vec{b}=(11100****)$\\
S-duplication
\end{center}
\end{minipage}

\begin{minipage}[t]{0.24\textwidth}
\begin{center}
\tikzstyle{vertexG}=[draw,circle,thick,fill=gray!50,minimum size=15pt]
\tikzstyle{vertexW}=[draw,circle,thick,fill=white,minimum size=15pt]
\tikzstyle{edge} = [draw,thick,->]

\begin{tikzpicture}[scale=0.75]
	\node[vertexG] (s) at (-0.75,-2) {$s$};
	\node[vertexW] (t) at (0.5,-1) {$t$};
	\path[edge] (-0.75,0) -- node[anchor=east] {$e_s$} (s);
	\path[edge] (0.5,0) -- node[anchor=west] {$e_t$} (t);
	\path[edge] (t) -- (0,-2);
	\path[edge] (t) -- (1,-2);
	\node[anchor=north] (s1) at (0,-2) {$e_{t_1}$};
	\node[anchor=north] (s2) at (1,-2) {$e_{t_2}$};
\end{tikzpicture}
$\vec{b}=(00000****)$\\
Null event
\end{center}
\end{minipage}
\begin{minipage}[t]{0.24\textwidth}
\begin{center}
\tikzstyle{vertexG}=[draw,circle,thick,fill=gray!50,minimum size=15pt]
\tikzstyle{vertexW}=[draw,circle,thick,fill=white,minimum size=15pt]
\tikzstyle{edge} = [draw,thick,->]

\begin{tikzpicture}[scale=0.75]
	\draw[pattern=north west lines] (-0.5,-1.5) -- (-0.25,-1.75) -- (-0.5,-2) -- (-1,-2) -- (-0.5,-1.5);
	\node[vertexW] (t) at (0.5,-1) {$t$};
	\node[vertexG] (s) at (-1,-2) {$s$};
	\path[edge] (-1.25,-1.25) -- node[anchor=south] {$e_s$} (-0.5,-1.5) -- (s);
	\path[edge] (0.5,0) -- node[anchor=west] {$e_t$} (t);
	\path[edge] (t) -- (-0.5,-2);
	\path[edge] (t) -- (1.5,-2);
	\node[anchor=north] (s1) at (-0.5,-2) {$e_{t_1}$};
	\node[anchor=north] (s2) at (1.5,-2) {$e_{t_2}$};
\end{tikzpicture}
$\vec{b}=(00010****)$\\
Landing
\end{center}
\end{minipage}
\begin{minipage}[t]{0.24\textwidth}
\begin{center}
\tikzstyle{vertexG}=[draw,circle,thick,fill=gray!50,minimum size=15pt]
\tikzstyle{vertexW}=[draw,circle,thick,fill=white,minimum size=15pt]
\tikzstyle{edge} = [draw,thick,->]

\begin{tikzpicture}[scale=0.75]
	\draw[pattern=north east lines] (0.5,-1.5) -- (0.75,-1.25) -- (1.5,-2) -- (1,-2) -- (0.5,-1.5);
	\node[vertexW] (t) at (0.5,-1) {$t$};
	\node[vertexG] (s) at (1,-2) {$s$};
	\path[edge] (-0.5,-1.25) -- node[anchor=south,pos=0.25] {$e_s$} (0.5,-1.5) -- (s);
	\path[edge] (0.5,0) -- node[anchor=west] {$e_t$} (t);
	\path[edge] (t) -- (-0.5,-2);
	\path[edge] (t) -- (1.5,-2);
	\node[anchor=north] (s1) at (-0.5,-2) {$e_{t_1}$};
	\node[anchor=north] (s2) at (1.5,-2) {$e_{t_2}$};
\end{tikzpicture}
$\vec{b}=(00001****)$\\
Landing
\end{center}
\end{minipage}
\begin{minipage}[t]{0.24\textwidth}
\begin{center}
\tikzstyle{vertexG}=[draw,ellipse,thick,fill=gray!50,minimum width=30pt]
\tikzstyle{vertexW}=[draw,circle,thick,fill=white,minimum size=15pt]
\tikzstyle{edge} = [draw,thick,->]

\begin{tikzpicture}[scale=0.75]
	\draw[fill=gray!50] (0,-1.5) -- (-0.5,-2) -- (0.5,-2) -- (0,-1.5);
 	\draw[pattern=north west lines] (-0.25,-1.25) -- (0,-1.5) -- (-0.5,-2) -- (-1,-2) -- (-0.25,-1.25);
	\draw[pattern=north east lines] (0,-1.5) -- (0.25,-1.25) -- (1,-2) -- (0.5,-2) -- (0,-1.5);
	\node[vertexW] (t) at (0,-1) {$t$};
	\draw[thick] (-1,-1.25) -- node[anchor=south,pos=0.25] {$e_s$} (0,-1.5);
	\path[edge] (0,-1.5) -- (-0.4,-1.9);
	\path[edge] (0,-1.5) -- (0.4,-1.9);
	\node[vertexG] (s) at (0,-2.1) {$s$};
	\path[edge] (0,0) -- node[anchor=west] {$e_t$} (t);
	\path[edge] (t) -- (-1,-2);
	\path[edge] (t) -- (1,-2);
	\node[anchor=north] (s1) at (-1,-2) {$e_{t_1}$};
	\node[anchor=north] (s2) at (1,-2) {$e_{t_2}$};
\end{tikzpicture}
$\vec{b}=(00011****)$\\
Double landing
\end{center}
\end{minipage}
\begin{minipage}[t]{0.24\textwidth}
\begin{center}
\tikzstyle{vertexG}=[draw,circle,thick,fill=gray!50,minimum size=15pt]
\tikzstyle{vertexW}=[draw,circle,thick,fill=white,minimum size=15pt]
\tikzstyle{edge} = [draw,thick,->]

\begin{tikzpicture}[scale=0.75]
	\draw[pattern=horizontal lines] (0,0) -- (0.5,0) -- (0.5,-0.5) -- (0,-0.5) -- (0,0);
	\node[vertexW] (t) at (0.5,-1) {$t$};
	\path[edge] (0,0) -- node[anchor=east] {$e_s$} (0,-0.5);
	\node (x) at (0,-0.5) {$\times$};
	\path[edge] (0.5,0) -- node[anchor=west] {$e_t$} (t);
	\path[edge] (t) -- (-0.5,-2);
	\path[edge] (t) -- (1.5,-2);
	\node[anchor=north] (s1) at (-0.5,-2) {$e_{t_1}$};
	\node[anchor=north] (s2) at (1.5,-2) {$e_{t_2}$};
\end{tikzpicture}
$\vec{b}=(100000000)$\\
Extinction / Association loss
\end{center}
\end{minipage}
\begin{minipage}[t]{0.24\textwidth}
\begin{center}
\tikzstyle{vertexG}=[draw,circle,thick,fill=gray!50,minimum size=15pt]
\tikzstyle{vertexW}=[draw,circle,thick,fill=white,minimum size=15pt]
\tikzstyle{edge} = [draw,thick,->]

\begin{tikzpicture}[scale=0.75]
	\draw[pattern=horizontal lines] (0,0) -- (0.5,0) -- (0.5,-1) -- (-0.5,-2) -- (-1,-2) -- (0,-1) -- (0,0);
	\node[vertexW] (t) at (0.5,-1) {$t$};
	\node[vertexG] (s) at (-1,-2) {$s$};
	\path[edge] (0,0) -- node[anchor=east] {$e_s$} (0,-1) -- (s);
	\path[edge] (0.5,0) -- node[anchor=west] {$e_t$} (t);
	\path[edge] (t) -- (-0.5,-2);
	\path[edge] (t) -- (1.5,-2);
	\node[anchor=north] (s1) at (-0.5,-2) {$e_{t_1}$};
	\node[anchor=north] (s2) at (1.5,-2) {$e_{t_2}$};
\end{tikzpicture}
$\vec{b}=(10010****)$\\
T-takeoff (S-sorting)
\end{center}
\end{minipage}
\begin{minipage}[t]{0.24\textwidth}
\begin{center}
\tikzstyle{vertexG}=[draw,circle,thick,fill=gray!50,minimum size=15pt]
\tikzstyle{vertexW}=[draw,circle,thick,fill=white,minimum size=15pt]
\tikzstyle{edge} = [draw,thick,->]

\begin{tikzpicture}[scale=0.75]
	\draw[pattern=horizontal lines] (0,0) -- (0.5,0) -- (0.5,-1) -- (1.5,-2) -- (1,-2) -- (0,-1) -- (0,0);
	\node[vertexW] (t) at (0.5,-1) {$t$};
	\node[vertexG] (s) at (1,-2) {$s$};
	\path[edge] (0,0) -- node[anchor=east] {$e_s$} (0,-1) -- (s);
	\path[edge] (0.5,0) -- node[anchor=west] {$e_t$} (t);
	\path[edge] (t) -- (-0.5,-2);
	\path[edge] (t) -- (1.5,-2);
	\node[anchor=north] (s1) at (-0.5,-2) {$e_{t_1}$};
	\node[anchor=north] (s2) at (1.5,-2) {$e_{t_2}$};
\end{tikzpicture}
$\vec{b}=(10001****)$\\
T-takeoff (S-sorting)
\end{center}
\end{minipage}
\begin{minipage}[t]{0.24\textwidth}
\begin{center}
\tikzstyle{vertexG}=[draw,circle,thick,fill=gray!50,minimum size=15pt]
\tikzstyle{vertexW}=[draw,circle,thick,fill=white,minimum size=15pt]
\tikzstyle{edge} = [draw,thick,->]

\begin{tikzpicture}[scale=0.75]
	\draw[pattern=horizontal lines] (0,0) -- (0.75,0) -- (0.75,-1) -- (1,-1) -- (1,-2) -- (0,-2) -- (0,0);
	%\draw[pattern=horizontal lines] (0,0) -- (0.75,0) -- (0.75,-1) -- (0,-1) -- (0,0);
	%\draw[pattern=north west lines] (0,-1) -- (1,-1) -- (1,-2) -- (0,-2) -- (0,-1);
	%\draw[pattern=north east lines] (0,-1) -- (0.5,-1) -- (0.5,-2) -- (0,-2) -- (0,0);
	\path[edge] (0.5,-1) -- (0.5,-2);
	\path[edge] (1,-1) -- (1,-2);
	\node[vertexG] (s) at (0,-2) {$s$};
	\node[vertexW] (t) at (0.75,-1) {$t$};
	\path[edge] (0,0) -- node[anchor=east] {$e_s$} (s);
	\path[edge] (0.75,0) -- node[anchor=west] {$e_t$} (t);
 	\node[anchor=north] (t1) at (0.5,-2) {$e_{t_1}$};
 	\node[anchor=north] (t2) at (1,-2) {$e_{t_2}$};
\end{tikzpicture}
$\vec{b}=(10011****)$\\
T-duplication
\end{center}
\end{minipage}
\end{center}
\end{scriptsize}
\caption{Coevolutionary events for case i), i.e., speciation of $s$ occurs before the speciation of $t$, are shown in the upper two rows. Events for case ii), i.e., speciation of $s$ occurs after the speciation of $t$, are shown in the lower two rows. A '*' in the association tuples represents an arbitrary binary value.}
\label{fig:legeEvents}
\end{figure}
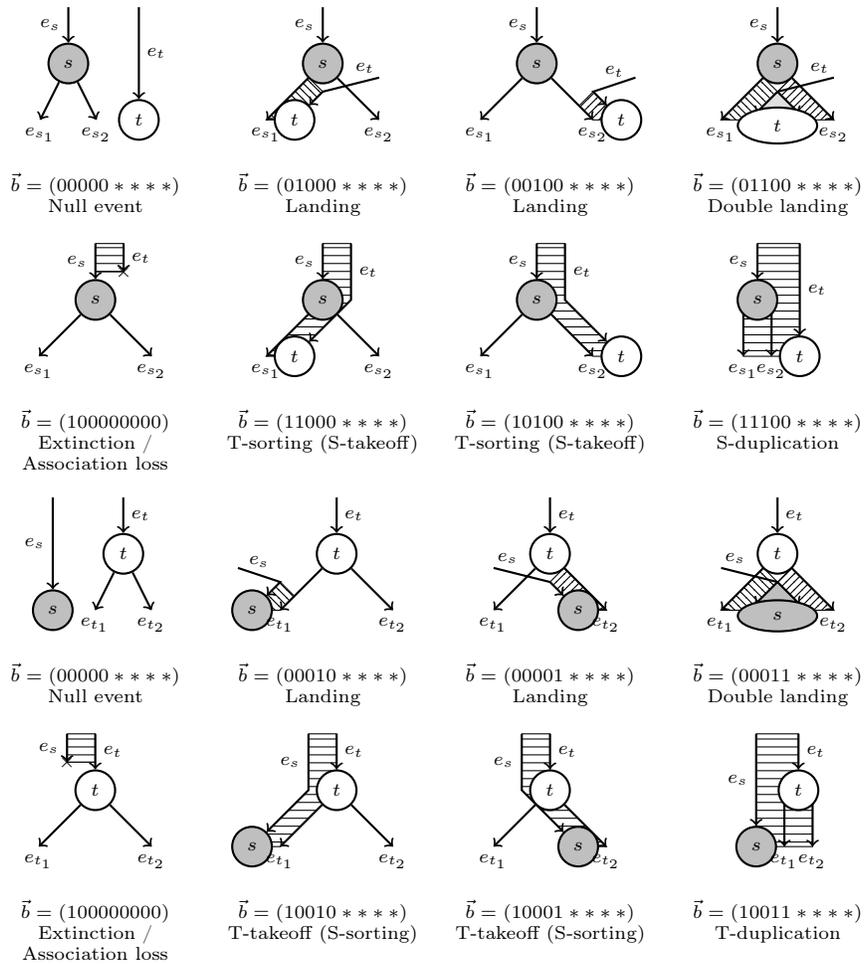

In the algorithmic sections the restricted set of phylogenetically meaningful events is considered only.

\subsection{Dynamic Programming}
\label{sec:dp}
For a given X-tanglegram $(S,T,\phi)$ and cost model $\gamma$ a reconciliation $\mathcal{R}$ is sought, such that the sum of all event and leaf-to-leaf association costs is minimal.

In the first step dynamic programming is used to determine all optimal sub-reconciliations with respect to a given cost model.
Therefore, two dynamic programming matrices $C^0$ and $C^1$ are computed. $C^0(s,t)$ gives the costs for the optimal sub-reconciliation $\mathcal{R}_{|s,t}$ of the two subtrees rooted at $s$ and $t$ with $s$ and $t$ being not associated. Accordingly, $C^1(s,t)$ computes the optimal sub-reconciliation costs with $s$ and $t$ being associated. Starting from a pair of extant species $(s,t)$ the cost of a (non-)association $C^b(s,t)$ is evaluated based on the a priori association strength $\phi(s,t)$.
For pairs of ancestral species $(s,t)$ the cost of each possible cophylogenetic event is evaluated, weighted, and accumulated with the costs of the respective sub-reconciliations. From all those costs the minimum is chosen for $C^b(s,t)$.
With $b=b_{s,t} \in \{0,1\}$ the dynamic programming formulation is as follows.
\begin{equation}
	C^b(s,t) =
	\left\{
	\begin{array}{ll}
		%g^b(\phi (s,t))-\beta                       & \textrm{if $s \in L_S$ and $t \in L_T$} \\
		\alpha(1/\phi(s,t)-1)                   & \textrm{if $s \in L_S$, $t \in L_T$ and $b=1$} \\
		\alpha(1/(1-\phi(s,t))-1)                & \textrm{if $s \in L_S$, $t \in L_T$ and $b=0$} \\
		C^b_<(s,t)                                    & \textrm{if $s \notin L_S$ and $t \in L_T$} \\
		C^b_>(s,t)                                    & \textrm{if $s \in L_S$ and $t \notin L_T$} \\
		\min (C^b_<(s,t), C^b_>(s,t), C^b_=(s,t))    & \textrm{otherwise}
	\end{array}
	\right.
	\label{eqn:costs}
\end{equation}
with the user defined parameter $\alpha \geq 1$ resulting in a cost value between $0$ and $\infty$ for the leaf-to-leaf associations (division by $0$ is evaluated as $\infty$). $C^b_\odot(s,t), \odot \in \{<, >, =\}$ is the cost of the minimal sub-reconciliation with an event of type $\odot$ occurring during the speciation of $s$ and/or $t$.
Precisely:

\begin{equation}
\begin{array}{lll}
C^b_<(s,t) & = & \min\limits_{b_{s_1,t},b_{s_2,t} \in \{0,1\}} \Big(\big(\sum\limits_{i \in \{1,2\}} C^{b_{s_i,t}}(s_i,t)\big) + g^b\big(\phi (s,t)\big)\cdot\gamma(\vec{b}_{s,t}) \Big)\\
C^b_>(s,t) & = & \min\limits_{b_{s,t_1},b_{s,t_2} \in \{0,1\}} \Big(\big(\sum\limits_{j \in \{1,2\}} C^{b_{s,t_j}}(s,t_j)\big) + g^b\big(\phi (s,t)\big)\cdot\gamma(\vec{b}_{s,t}) \Big)\\
C^b_=(s,t) & = & \min\limits_{\substack{b_{s_x,t_y} \in \{0,1\},\\x,y \in \{1,2\}}} \Big(\big(\sum\limits_{i,j \in \{1,2\}} C^{b_{s_i,t_j}}(s_i,t_j)\big) + g^b\big(\phi (s,t)\big)\cdot\gamma(\vec{b}_{s,t}) \Big)
\end{array}
\end{equation}
with 
$g^0(x) =
\left\{
\begin{array}{ll}
	\infty                & \text{if } x = 1\\
	\beta^{(2x-1)}       & \text{otherwise}\\
\end{array}
\right.$
and
$g^1(x) =
\left\{
\begin{array}{ll}
	\infty                & \text{if } x = 0\\
	\beta^{-(2x-1)}       & \text{otherwise}\\
\end{array}
\right.$
.

The functions $g^b(x), b \in \{0,1\}$ give the weighting factors for the event costs $\gamma(\vec{b}_{s,t})$.
The user defined parameter $\beta \geq 1$ results in factors ranging from $1/\beta$ to $\beta$. Associations with a strength of $x=0$, respectively non-associations with a strength of $x=1$, result in infinite costs. The term $(2x-1)$ is used to normalize $g^b(x)$ to result in a factor of $1$ if the association strength is $0.5$.

After computing the dynamic programming matrices the cost for an optimal reconciliation is given by $\min(C^0_{\rho_S,\rho_T},C^1_{\rho_S,\rho_T})$ with $\rho_S$ and $\rho_T$ being the roots of trees $S$ and $T$. The reconciliation can be retrieved by backtracking.

\subsection{Time Consistency}
\label{sec:time}
\begin{figure}
\begin{center}
% \floatbox[{\capbeside\thisfloatsetup{capbesideposition={right,top},capbesidewidth=0.40\textwidth}}]{figure}[\FBwidth]
% {\caption{\textbf{(a)} The association between the two edges $e_s$ and $e_t$ results in the two timing constraints $\tau(s')<\tau(t)$ and $\tau(t')<\tau(s')$ (dashed arrows). \textbf{(b)} The four associations $(e_{s1},e_{t1})$, $(e_{s2},e_{t2})$, $(e_{s1'},e_{t2_j})$ and $(e_{s2'},e_{t1_i})$ result in circular timing constrains (cycle of dashed arrows).}\label{fig:timing}}
{\begin{minipage}[t]{0.20\textwidth}
\begin{center}
\tikzstyle{vertexG}=[draw,circle,thick,fill=gray!50,minimum size=20pt]
\tikzstyle{vertexW}=[draw,circle,thick,fill=white,minimum size=20pt]
\tikzstyle{edge} = [draw,thick,->]

\begin{tikzpicture}[scale=0.75, every node/.style={scale=0.75}]
	\draw[pattern=horizontal lines] (0,0) -- (0,-2.5) -- (1,-2.5) -- (1,0) -- (0,0);
	\node[vertexG] (ps) at (0,0) {$s'$};
	\node[vertexW] (pt) at (1,0) {$t'$};
	\node[vertexG] (s) at (0,-2.5) {$s$};
	\node[vertexW] (t) at (1,-2.5) {$t$};
	\path[edge] (0,1) -- (ps);
	\path[edge] (1,1) -- (pt);
	\path[edge] (ps) -- node[anchor=east] {$e_s$} (s);
	\path[edge] (pt) -- node[anchor=west] {$e_t$} (t);
	\path[edge,dashed,line width=2pt] (ps) -- (t);
	\path[edge,dashed,line width=2pt] (pt) -- (s);
	\path[edge] (s) -- (0,-3.5);
	\path[edge] (t) -- (1,-3.5);
\end{tikzpicture}\\
(a)
\end{center}
\end{minipage}
\begin{minipage}[t]{0.40\textwidth}
\begin{center}
\tikzstyle{vertexG}=[draw,circle,thick,fill=gray!50,minimum size=15pt]
\tikzstyle{vertexW}=[draw,circle,thick,fill=white,minimum size=15pt]
\tikzstyle{edge} = [draw,thick,->]

\begin{tikzpicture}[scale=0.575, every node/.style={scale=0.75}]
	\draw[pattern=horizontal lines] (-1,-1) -- (-0.5,-1) -- (-1.5,-2) -- (-2,-2) -- (-1,-1);
	\draw[pattern=horizontal lines] (-2.5,-2) -- (-2,-2) -- (-2,-4) -- (-2.5,-4) -- (-2.5,-2);
	\draw[pattern=horizontal lines] (0.5,-1) -- (1,-1) -- (2,-2) -- (1.5,-2) -- (0.5,-1);
	\draw[pattern=horizontal lines] (2,-2) -- (2.5,-2) -- (2.5,-4) -- (2,-4) -- (2,-2);
	\node[vertexG] (rhoS) at (0,0) {$\rho_S$};
	\node[vertexW] (rhoT) at (0,1) {$\rho_T$};
	\node[vertexG] (ps1) at (-2,-2) {$s1'$};
	\node[vertexG] (ps2) at (2,-2) {$s2'$};
	\draw[thick,dashed] (rhoS) -- (-1,-1);
	\draw[thick,dashed] (rhoS) -- (1,-1);
	\path[edge] (-1,-1) -- node[anchor=east] {$e_{s1'}$} (ps1);
	\path[edge] (1,-1) -- node[anchor=west] {$e_{s2'}$} (ps2);
	\path[edge] (ps1) -- node[anchor=west,pos=0.25] {$e_{s1}$} (-2,-4);
	\path[edge] (ps2) -- node[anchor=east,pos=0.25] {$e_{s2}$} (2,-4);
	\node[vertexW] (t1) at (-2.5,-4) {$t1$};
	\node[vertexW] (t2) at (2.5,-4) {$t2$};
	\draw[thick,dashed] (rhoT) -- (-2.5,0) -- (-2.5,-2);
	\draw[thick,dashed] (rhoT) -- (2.5,0) -- (2.5,-2);
	\path[edge] (-2.5,-2) -- node[anchor=east] {$e_{t1}$} (t1);
	\path[edge] (2.5,-2) -- node[anchor=west] {$e_{t2}$} (t2);
	\path[edge] (t1) -- (0.5,-1) -- node[anchor=east] {$e_{t1_i}$} (1.5,-2);
	\path[edge] (t2) -- (-0.5,-1) -- node[anchor=west] {$e_{t2_j}$} (-1.5,-2);
	\path[edge,dashed,line width=2pt] (ps1) -- (t1);
	\path[edge,dashed,line width=2pt] (t1) -- (ps2);
	\path[edge,dashed,line width=2pt] (ps2) -- (t2);
	\path[edge,dashed,line width=2pt] (t2) -- (ps1);
\end{tikzpicture}\\
(b)
\end{center}
\end{minipage}}
\end{center}
\caption{\textbf{(a)} The association between the two edges $e_s$ and $e_t$ results in the two timing constraints $\tau(s')<\tau(t)$ and $\tau(t')<\tau(s')$ (dashed arrows). \textbf{(b)} The four associations $(e_{s1},e_{t1})$, $(e_{s2},e_{t2})$, $(e_{s1'},e_{t2_j})$ and $(e_{s2'},e_{t1_i})$ result in circular timing constrains (cycle of dashed arrows).}\label{fig:timing}
\end{figure}
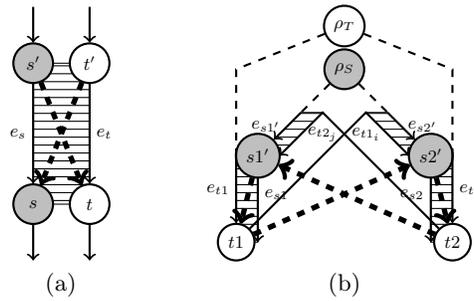

Although the dynamic programming solutions are optimal with respect to the given cost model $\gamma$, the reconciliations might be phylogenetically invalid due to chronological inconsistencies \cite{Merkle:05}. With the dynamic programming formulation it is assured that for an association of species $s$ and $t$ no descendant of $s$ is associated with an ancestor of $t$, and vice versa. However, additional timing constraints are introduced by each pairwise association in $\mathcal{R}$. Assume two species $s$ and $t$ being associated and therefore $(e_s,e_t) \in \mathcal{R}$. As species $s$ and $t$ interacted, both existed at the same time. Hence, species $s$ has to be emerged before the speciation of $t$ and vice versa and it follows that $\forall s' <_T s : \tau(s')<\tau(t)$, respectively $\forall t' <_T t : \tau(t')<\tau(s)$, see Figure \ref{fig:timing} (a).
This is not assured by the dynamic programming formulation. Contradicting associations of species from disjoint subtrees of both phylogenies can occur. For an example consider the species $s1$, $s2$, $t1$, and $t2$ from disjoint subtrees of $S$, respectively $T$, with $\{(e_{s1},e_{t1}),(e_{s2},e_{t2})\} \subseteq \mathcal{R}$. Furthermore, let the species $s1'$ and $s2'$ be the parent species of $s1$, respectively $s2$. Now assume that $s1'$ is associated with the child species $t2_j$ of $t2$ and $s2'$ is associated with the child species $t1_i$ of $t1$, i.e., $\{(e_{s1'},e_{t2_j}),(e_{s2'},e_{t1_i})\} \subseteq \mathcal{R}$. This scenario creates (among others) the timing constraints $\tau(s1')<\tau(t1)<\tau(s2')<\tau(t2)<\tau(s1')$ which gives a timing inconsistency. Figure \ref{fig:timing} (b) shows this example.
For a reconciliation to be phylogenetically meaningful such cases must not occur. To determine phylogenetically valid reconciliations the following two definitions are needed.

\begin{definition}[Timing Graph]
For a given X-tanglegram $(S,T,\phi)$ and a reconciliation $\mathcal{R}$ the \textit{timing graph} TG is a directed graph $(V_{TG},E_{TG})$ with node set $V_{TG}=V_S \cup V_T$ and edge set  $E_{TG}=E_S \cup E_T \cup \{(s',t),(t',s) : (e_s,e_t) \in \mathcal{R}$ and $(s',s) \in E_S$ and $(t',t) \in E_T\}$.
\end{definition}
 
\begin{definition}[Time Consistency]
A reconciliation $\mathcal{R}$ for an X-tanglegram $(S,T,\phi)$ is said to be time consistent if its \textit{timing graph} is acyclic.
\end{definition}

Observe that there are event models where the dynamic programming will always create time consistent reconciliations which are parsimonious with respect to the total event costs.
Chronological inconsistencies are created by contradicting timing constraints. But the set of timing constraints coming from the tree topologies of both trees without any associations are always compatible. Therefore, for an inconsistency to occur there must be paths in the timing graph connecting nodes of $S$, respectively $T$, which have no ancestor/descendant relation, i.e., nodes from distinct subtrees of the same phylogenetic tree. Such constraints are created by either S-duplications (only between nodes from $S$) and T-duplications (only between nodes from $T$) or by any of the landing events (constraints between nodes from both trees). A chronological inconsistency can be created only if combinations of such constrains between both types of trees exist. Therefore, any event model which forbids landing events and allows only one type of duplication (S or T) will lead to time consistent reconciliations. For instance the gene duplication/gene loss event model commonly used for gene tree/species tree reconciliation can be solved without chronological inconsistencies using dynamic programming.

But even if events are considered which possibly lead to chronological inconsistencies, dynamic programming can be used to construct valid solutions as well. This is done by running dynamic programming and checking the timing graph for cycles. In each cycle there exist edges which were introduced due to associations between species. In an iterative way one can restrict the input data by forbidding some of the associations and redo the dynamic programming until the timing graph is acyclic. However, the produced reconciliation may not be parsimonious anymore, see also \cite{Patro:12}.

\subsection{Branch-and-Bound Algorithm}
\label{sec:bnb}

Although the (unrestricted) dynamic programming does not necessarily result in a time consistent reconciliation, it can be used to determine a lower bound for the costs of a partially computed reconciliation. This lower bound is used by a branch-and-bound algorithm to cut the computation whenever a partial solution indicates that the cost will be higher than a certain value, e.g., the cost of a previously found reconciliation or a maximum threshold.

The algorithm starts with an empty set of associations and a lower bound of $\min(C^0_{\rho_S,\rho_T},C^1_{\rho_S,\rho_T})$ for the reconciliation costs.
In the branching procedure of the algorithm a decision tree is traversed by considering in each step a pair of species, i.e., edges, as either associated or not associated. In this way a partial reconciliation is constructed adding at most one association per branching step.
The pairs are selected in a top-down manner, starting from $(\rho_S,\rho_T)$, such that it is assured that when selecting a pair $(s,t)$ all pairs of ancestral species $(s',t')$, with $s' \prec_S s$ and $t' \prec_T t$, were already processed beforehand.
For each selection the timing graph is updated and checked for cycles. If the graph contains any cycles the computation is cut at this point and the next pair of species is selected.

As a cophylogenetic event consists of multiple associations, all the respective pairs have to be processed before an event, and therefore its cost, can be determined. If this is the case then the cost of the respective event is added to the cost of the already computed partial reconciliation.
The lower bound for the total costs is then given by the costs of the partial reconciliation plus the minimum costs needed for the sub-reconciliations of the unprocessed pairs of subtrees. These minimum costs are taken from the two dynamic programming matrices $C^0$ and $C^1$.

To avoid unnecessary branching after choosing two species $s$ and $t$ as being associated, all pairs $(s',t'')$ and $(s'',t')$ with $s' \prec_S s \prec_S s''$ and $t' \prec_T t \prec_T t''$ are assumed to be not associated, as this would result in timing inconsistencies. The pseudocode is given in Algorithm \ref{alg:bnb}.

Beside timing constraints, some coevolutionary systems require that additional properties have to be satisfied for a reconciliation to be valid.
There may be systems which require, e.g., that each species of one type is associated with at least, respectively at most, one species of the other type at a time.
For instance in gene tree/species tree reconciliations each (ancestral) gene is associated with exactly one species.
As these constraints restrict the set of valid solutions they can be easily integrated into the branch-and-bound algorithm. In each branching step it is checked if the current association will lead to a violation and, if applicable, the computation can be cut.

\begin{algorithm}
\KwIn{$pairs$: the queue of unprocessed pairs - $[(\rho_S,\rho_T)]$ at first call\;
$A[][]$: the association matrix; $cost_{partial}$: the costs of the  partial reconciliation\; 
$cost_{bounds}$: the lower bound for the costs of unprocessed sub-reconciliations\;
$cost_{best}$: cost of current best solution or infinity/max threshold\;
$\text{TG}$: the timing graph; the trees $S$ and $T$ and DP matrices $C^0$ and $C^1$;}
%\KwOut{Minimal sum of event costs for the cophylogenetic reconciliation of $S$ and $T$}

\If{$pairs$ is not empty}{
	get next unprocessed pair $(s,t)$ from $pairs$\;
	add all pairs $(s,t_j),(s_i,t),(s_i,t_j)$ with $i,j \in \{1,2\}$ to $pairs$\;
	\ForEach{$b$ in $\{$``not-associated'', ``associated''$\}$}{
		\If{$A[s][t]$ is undefined or equals $b$}{
			set $A[s][t]$ to $b$\;
			\If{$b$ equals ``associated''}{
				\ForEach{$s'$, $s''$, $t'$, $t''$ with $s' <_T s <_T s''$ and $t' <_T t <_T t''$}{
					$A[s'][t''] \leftarrow$ ``not-associated''; $A[s''][t'] \leftarrow$ ``not-associated''\;
				}
			}
			update timing graph $\text{TG}$\;
			\If{timing is consistent}{
				\If{a new event can be determined from $A[][]$}{
					update $cost_{partial}$; update $cost_{bounds}$\;
				}
				\If{$cost_{partial}+cost_{bounds} \leq cost_{best}$}{
					$cost_{total} \leftarrow$ result of recursive call to Alg. \ref{alg:bnb}\;
					$cost_{best} \leftarrow \min(cost_{best},cost_{total})$\;
				}
			}
			undo changes to $\text{TG}$, $A[][]$, $cost_{partial}$, and $cost_{bounds}$\;
		}
	}
}
\Return $cost_{best}$
\caption{Pseudocode for computing the minimal sum of event costs for the cophylogenetic reconciliation of $S$ and $T$}
\label{alg:bnb}
\end{algorithm}

\section{Discussion}
\label{sec:discussion}
%\subsubsection{Event model} 
A certain type of application usually requires only a subset of the events modeled by this approach. For each event from this subset a cost value has to be specified. All neglected events get infinite costs and will therefore not occur in reconciliations with finite overall costs.
Obviously there is a 1-to-1 correspondence  between most of the cophylogenetic events depicted in Table \ref{tab:events} and the events defined by local association patterns shown in Figures \ref{fig:legeEvents} and \ref{fig:eqEvents}. Only the two types of \emph{switches} (partial and complete) and the \emph{lateral transfer from the dead} can not be modeled directly by this approach. Instead, a \emph{partial switch} is seen as a combination of the events \emph{takeoff} and \emph{landing}, while a \emph{complete switch} and a \emph{lateral transfer from the dead} are modeled by an \emph{extinction} and a \emph{landing}. When giving the \emph{takeoff} and \emph{landing} events, respectively \emph{extinction} and \emph{landing} events, the same accumulated costs, then a certain reconciliation results in the same overall costs in both event models.
This opens up new possibilities for solving reconciliation problems for a variety of applications. Biogeography, gene tree/species tree, and host-parasite systems can be reconciled with the same algorithms while only the cost model $\gamma$ differs. Beside that, further cases of application exists, e.g., general symbiotic systems or interactions of genes or gene products, where both association partners are equitable and a reconciliation can not be produced by simply embedding one tree into the other.

%\subsubsection{Edge mapping vs. node mapping}
In this approach we decided a reconciliation to be a set of associations between edges. This has been done to retain a similar graphical representation for the cophylogenetic events as it has been used in previous publications. However, as more complex events can be modeled by this approach, e.g., the \emph{partial cospeciation}, visualizing a reconciliation is a complex task and the interpretation of the results might be hard. An alternative is to redefine a reconciliation to be a set of associations between nodes on the directed line graphs (line digraphs) of the phylogenetic trees with artificial root $\rho'$. Both definitions are interchangeable as in the line graph only the roles of nodes and edges are changed but the tree structure is retained. Then a reconciliation can be depicted as a graph with directed \emph{tree edges} representing the tree topology of both trees and undirected \emph{association edges}.
 
%\subsubsection{Complexity} 
The dynamic programming algorithm has a time complexity of $O(n^2)$, where $n$ is the number of leaves in the larger phylogenetic tree. For each of the $O(n^2)$ many values from the two matrices $C^b_<(s,t)$, $b\in\{0,1\}$, the cost of a constant number of possible association patterns is determined. For each pattern this can be done in constant time.

\section{Conclusion}
\label{sec:conclusion}
In this paper we introduced a cophylogenetic event model based on local association patterns between coincident edges. Due to the large variety of possible events and the possibility to neglect events via a corresponding cost model the approach suits to various reconciliation problems. In addition we provided the possibility to use a priori knowledge about association strengths to improve the reconciliations. We presented a $O(n^2)$ time heuristic based on dynamic programming as well as an exact branch-and-bound algorithm to solve the reconciliation problem for a given X-tanglegram $(S,T,\phi)$ and cost model $\gamma$. 

Until now, only binary phylogenetic trees are considered. But extending the approach to support polytomies can be done by extending the event model to non-binary events. However, this requires a fixed maximal outdegree for the internal nodes and therefore can not be used for multifurcations of arbitrary size. Alternatively, the polytomies can be resolved within the reconciliation process using heuristic approaches.

In recent years there is a trend towards using also phylogenetic networks. These networks can display hybridization events, as nodes with indegree greater than one. While extending our event model to consider cophylogenetic hybridization is straightforward, further research is needed to adopt the reconciliation algorithms.

\section*{Acknowledgements}
This work was funded by the German Research Foundation (DFG) through the project MI439/14-1.

\bibliographystyle{splncs03}
\bibliography{references}

\begin{thebibliography}{10}
\providecommand{\url}[1]{\texttt{#1}}
\providecommand{\urlprefix}{URL }

\bibitem{Berglund:06}
Berglund-Sonnhammer, A.C., Steffansson, P., Betts, M., Liberles, D.: Optimal
  gene trees from sequences and species trees using a soft interpretation of
  parsimony. J Mol Evol  63(2),  240--250 (2006)

\bibitem{Brooks:90}
Brooks, D.: Parsimony analysis in historical biogeography and coevolution:
  methodological and theoretical update. Syst Biol  39(1),  14--30 (1990)

\bibitem{Charleston:98}
Charleston, M.: Jungles: a new solution to the host/parasite phylogeny
  reconciliation problem. Math Biosci  149(2),  191--223 (1998)

\bibitem{Charleston:06}
Charleston, M., Perkins, S.: Traversing the tangle: algorithms and applications
  for cophylogenetic studies. J Bio Med Inform  39(1),  62--71 (2006)

\bibitem{Chen:00}
Chen, K., Durand, D., Farach-Colton, M.: {NOTUNG}: a program for dating gene
  duplications and optimizing gene family trees. J Comput Biol  7(3-4),
  429--447 (2000)

\bibitem{Conow:10}
Conow, C., Fielder, D., Ovadia, Y., Libeskind-Hadas, R.: Jane: a new tool for
  the cophylogeny reconstruction problem. Algorithms Mol Biol  5, ~16 (2010)

\bibitem{Doyon:10}
Doyon, J.P., Scornavacca, C., Gorbunov, K.Y., Sz{\"o}llosi, G., Ranwez, V.,
  Berry, V.: An efficient algorithm for gene/species trees parsimonious
  reconciliation with losses, duplications and transfers (2010)

\bibitem{Goodman:79}
Goodman, M., Czelusniak, J., Moore, G., Romero-Herrera, A., Matsuda, G.:
  Fitting the gene lineage into its species lineage, a parsimony strategy
  illustrated by cladograms constructed from globin sequences. Syst Biol
  28(2),  132--163 (1979)

\bibitem{Hafner:88}
Hafner, M., Nadler, S.: Phylogenetic trees support the coevolution of parasites
  and their hosts. Nature  332(6161),  258--259 (1988)

\bibitem{Hafner:90}
Hafner, M., Nadler, S.: Cospeciation in host-parasite assemblages: comparative
  analysis of rates of evolution and timing of cospeciation events. Syst Biol
  39(3),  192--204 (1990)

\bibitem{Hendy:84}
Hendy, M., Little, C., Penny, D.: Comparing trees with pendant vertices
  labelled. Siam J Appl Math  44(5),  1054--1065 (1984)

\bibitem{Merkle:05}
Merkle, D., Middendorf, M.: Reconstruction of the cophylogenetic history of
  related phylogenetic trees with divergence timing information. Theory Biosci
  123(4),  277--299 (2005)

\bibitem{Merkle:10}
Merkle, D., Middendorf, M., Wieseke, N.: A parameter-adaptive dynamic
  programming approach for inferring cophylogenies. BMC Bioinformatics
  11(S-1), ~60 (2010)

\bibitem{Nelson:91}
Nelson, G., Ladiges, P.: Three area statements: standard assumptions for
  biogeographic analysis. Syst Zool  40,  470--485 (1991)

\bibitem{Nelson:81}
Nelson, G., Platnick, N.: Systematics and biogeography: cladistics and
  vicariance. Columbia University Press (1981)

\bibitem{Ovadia:11}
Ovadia, Y., Fielder, D., Conow, C., Libeskind-Hadas, R.: The co-phylogeny
  reconstruction problem is {NP}-complete. J Comput Biol  18(1),  59--65 (2011)

\bibitem{Page:03}
Page, R.D.M.: Tangled Trees. Phylogeny, Cospeciation and Coevolution. The
  University of Chicago Press (2003)

\bibitem{Page:88}
Page, R.: Quantitative cladistic biogeography: constructing and comparing area
  cladograms. Soc Syst Zool (1988)

\bibitem{Page:94}
Page, R.: Maps between trees and cladistic analysis of historical associations
  among genes, organisms, and areas. Syst Biol  43(1),  58--77 (1994)

\bibitem{Page:98}
Page, R.: {GeneTree}: comparing gene and species phylogenies using reconciled
  trees. Bioinformatics  14(9),  819--820 (1998)

\bibitem{Page:98b}
Page, R., Charleston, M.: Trees within trees: phylogeny and historical
  associations. Trends Ecol Evol  13(9),  356--359 (1998)

\bibitem{Patro:12}
Patro, R., Sefer, E., Malin, J., Mar{\c{c}}ais, G., Navlakha, S., Kingsford,
  C.: Parsimonious reconstruction of network evolution. Algorithms Mol Biol
  7(1), ~25 (2012)

\bibitem{Ronquist:97}
Ronquist, F.: Dispersal-vicariance analysis: a new approach to the
  quantification of historical biogeography. Syst Biol  46(1),  195--203 (1997)

\bibitem{Ronquist:90}
Ronquist, F., Nylin, S.: Process and pattern in the evolution of species
  associations. Syst Biol  39(4),  323--344 (1990)

\bibitem{Ronquist:11}
Ronquist, F., Sanmartín, I.: Phylogenetic methods in biogeography. Annu Rev
  Ecol Evol Syst  42,  441--464 (2011)

\bibitem{Szollosi:13}
Szollosi, G., Tannier, E., Lartillot, N., Daubin, V.: Lateral gene transfer
  from the dead. Syst Biol  62(3),  386--397 (2013)

\bibitem{Vernot:08}
Vernot, B., Stolzer, M., Goldman, A., Durand, D.: Reconciliation with
  non-binary species trees. J Comput Biol  15(8),  981--1006 (2008)

\bibitem{Zandee:87}
Zandee, M., Roos, M.: Component-compatibility in historical biogeography.
  Cladistics  3,  305--332 (1987)

\bibitem{Zmasek:01}
Zmasek, C., Eddy, S.: A simple algorithm to infer gene duplication and
  speciation events on a gene tree. Bioinformatics  17(9),  821--828 (2001)

\end{thebibliography}

\end{document}